\documentclass[prd,aps,nofootinbib,showpacs]{revtex4}
\usepackage{amsmath}
\usepackage{amsmath,graphicx,color,epsfig}

\begin{document}
\title{Decays of $B$ meson to two charmed mesons}

\author{Run-Hui Li$^{1,2}$, Cai-Dian L\"u$^{1,4}$,  A.I. Sanda$^3$, and Xiao-Xia Wang$^1$}
\affiliation{
 \it $^1$ Institute of High Energy Physics, P.O. Box 918(4), Beijing 100049, People's Republic of China\\
 \it $^2$  School of Physics, Shandong University, Jinan 250100, People's Republic of China \\
$^3$Faculty of Technology, Kanagawa University, Yokohama, Kanagawa
221, Japan\\
 \it $^4$ Theoretical Physics Center for Science
Facilities, Beijing 100049, People's Republic of China}

\begin{abstract}
The factorization theorem in decays of $B_{(s)}$ mesons to two
charmed mesons (both pseudoscalar and vector) can be proved in the
leading order in  $m_D/m_B$ and $\Lambda_{\rm{QCD}}/m_D$ expansion.
Working in the perturbative QCD approach, we find  that the
factorizable emission diagrams are dominant. Most of branching
ratios we compute agree with the experimental data well, which means
that the factorization theorem seems to be reliable in predicting
branching ratios for these decays. In the decays of a $B$ meson to
two vector charmed mesons, the transverse polarization states
contribute $40\%-50\%$ both in the processes with an external W
emission and in the pure annihilation decays. This is in agreement
with the present experimental data. We also calculate the CP
asymmetry parameters. The results show that the direct CP
asymmetries are very small. Thus observation of any large direct CP
asymmetry will be a signal for new physics. The mixing induced CP
asymmetry in the neutral modes is large. This is also in agreement
with the current experimental measurements. They can give a cross
check of the $\sin 2\beta$ measurement from other channels.
\end{abstract}
 \maketitle

 \section{introduction}

The hadronic decays of B meson  are important for particle physics
since they provide constraints of the standard model
Cabibbo-Kobayashi-Maskawa (CKM) matrix, a test of the QCD
  factorization, information on the decay
mechanism, and the final state interaction. The CP asymmetries, in
which some of the hadronic uncertainties are canceled in their
theoretical predictions, play an important role in the
investigations of B physics. For the decays with a single D meson in
the final states, only tree operators contribute, and thus no CP
asymmetry appears in the standard model \cite{dpi}. However, for
decays with double-charm final states, there are penguin operator
contributions as well as tree operator contributions. Thus the
direct CP asymmetry may be present. Recently, the Belle
Collaboration reported a large direct CP violation in $B^0\to
D^+D^-$ decay~\cite{Belle:BtoDpDm}, while BaBar reproted a small
one, with a different sign even~\cite{Babar:BtoDpDm}. What is more,
large direct CP asymmetries have not been observed in other $B^0\to
D^{*+}D^{*-}$ decays~\cite{exp:BtoDD} either, which have the same
flavor structures as $B^0\to D^+D^-$ at the quark level. Intrigued
by these experimental results, many investigations on the decays of
B to double-charm states have been carried
out\cite{Fleischer:2007zn,Gronau:2008ed,Kim:2008ex,dsds}.

The theoretical study of hadronic B decays has achieved  great
success in recent years. Among them, the perturbative QCD approach
(PQCD) is based on $k_T$ factorization \cite{pqcd}. By keeping the
transverse momentum of quarks, the end point singularity in the
collinear factorization has been eliminated. Since transverse
momentum introduces another energy scale, double logarithm appears
in the QCD radiative corrections. The renormalization group equation
is used to resum the double logarithm, which results in the Sudakov
factor. This factor effectively suppresses the endpoint contribution
of the distribution amplitude of mesons in the small transverse
momentum region, which makes the perturbative calculation reliable.
Phenomenologically, the PQCD approach successfully predict the
following: (1)the direct CP asymmetry in B decays \cite{direct},
(2)the pure annihilation type B decays \cite{anni} (3) the strong
final state interaction phase and color suppressed decay amplitude
in the $B\to D \pi $ decays \cite{dpi}.

In charmless two-body B decays, the final state mesons can be
considered as massless therefore both of the final state mesons are
on the light cone. The collinear factorization can be easily proved
in the heavy quark limit. For the decays with a single heavy D meson
in the final states, one can still prove factorization \cite{scet}
in the leading order of the $r=m_D/m_B$ expansion. For the decays
with double-charm quarks in the final states, such as $B\to J/\psi
K$, $\chi_c K$, it is believed that the factorization fails. However
the decays with double D mesons in the final states are different.
The reason is that the expansion parameter $m_D/m_B\sim 0.36$ could
be considered small, $m_{J/\psi}/m_B\sim 0.6$ is not. In other
words, the $J/\psi $ ($\chi_c $) are soft particles in B decays;
while the $D_{(s)}^{(*)}$ meson is collinear in the $B \to
D_{(s)}^{(*)} D_{(s)}^{(*)}$ decays. The momentum of the
$D_{(s)}^{(*)}$  meson in the latter decays is $|\vec{p}|\simeq
\frac{1}{2}m_B (1-2r^2)$, which is  still nearly half of the B meson
mass. The decays of B to double-charm states can be investigated in
the PQCD approach in the leading order of $r=m_D/m_B$ and
$\Lambda_{\rm{QCD}}/m_D$ expansion. All of the annihilation type
diagrams contain end-point sigualrity,which are quite different from
the spectatorlike diagrams which are dominated by the form factors.
It is very difficult to deal with in the collinear factorization.
The PQCD base on $k_T$ factorization is almost the only approach
that can give quantitative calculations of annihilation type decays.

This paper is organized as follows. In Sec.~\ref{sec:analytic}, we
list the formalism, including the Hamiltonian, the wave functions of
the mesons, the factorization formulae of the Feynman diagrams for
$B\to PP$ decay mode, and the analytic expressions for the decay
amplitudes. In Sec.~\ref{sec:numerical}, the numerical results of
the physical observables and discussions of the results are given.
Sec.~\ref{Sec:summary} is a brief summary. The common PQCD
functions, scales, and the factorization formulae of the Feynman
diagrams for $B\to PV$, $B\to VP$, and $B\to VV$ modes are all put
into the appendices for simplicity.

\section{Analytic expressions}
\label{sec:analytic}

In hadronic B decays there are several typical energy scales, and
expansions with respect to the ratios of the scales are usually
carried out. The physics with a scale higher than the W boson mass
are electroweak interactions, which can be calculated
perturbatively. The physics between the W boson mass and b quark
mass obtain QCD corrections. This correction is included in the
Wilson coefficients of the four-quark operators in the effective
Hamiltonian. The physics below the b quark mass is more complicated.
We have to utilize the factorization theorem to factorize the
nonperturbative contributions out, so that the hard part can be
calculated perturbatively. In the PQCD approach, we utilize the
$k_T$ factorization \cite{pqcd}, where the transverse momenta of the
quarks in the mesons are kept to eliminate the end-point sigularity.
Because of the new transverse momentum scale introduction, double
logarithms appear in the calculation. We resum these logarithms to
give a Sudakov factor, which effectively suppresses the end-point
region contribution. Thus the end-point sigularity in the usual
collinear factorization disappears. This makes the perturbative
calculation reliable and consistent. For decays with D meson in the
final states, another scale $m_D$ is introduced. The factorization
is proved in the leading order of the $m_D/m_B$ expansion
\cite{scet}, therefore, as it is done in the computation of $B\to D
M$ and $B\to \bar D M$ amplitudes\cite{Li:2008ts}, we will work in
the leading order $m_D/m_B$ expansion. For each of the diagrams in
the following, we keep the contributions in the leading order of
$m_D/m_B$. For example, in the $B$ meson to two vector mesons
decays, the leading order contributions of some transversely
polarized amplitudes are proportional to $r^2$ ($r=m_D/m_B$). Then
we will keep the $r^2$ terms in these diagrams. While in other
cases, the terms of $r^2$ are neglected because the leading order is
lower than $2$. Finally the amplitude for $B\to M_2 M_3$ ($M_2$ and
$M_3$ stand for two mesons) decay within PQCD approach is decomposed
as
\begin{eqnarray}
{\cal M}= \int d^4{{k}}_{1}d^4{ {k}}_{2}d^4{ {k}}_{3}\Phi_B({
{k}}_{1},t) T_H({{k}}_{1},{{k}}_{2},{{k}}_{3},t) \Phi_{M_2}({
{k}}_{2},t)\Phi_{M_3}( { {k}}_{3},t) e^{S({k_i,t})},
\end{eqnarray}
where $k_i$ ($i=1,2,3$) are the momenta of the quarks in mesons
which are defined explicitly in Eq.(\ref{eq:fmoments}). $T_H$ is the
hard part that is perturbatively calculable. $\Phi_B$ and
$\Phi_{M_i}$ ($i=2,3$) are the hadronic meson wave functions that
are treated as nonperturbative inputs. The Sudakov factors
$e^{S({k_i,t})}$ ($i=1,2,3$) are from the resummation of double
logarithms .

\subsection{Notations and conventions}

The Hamiltonian referred to in this paper is given
by~\cite{Buchalla:1995vs}:
 \begin{eqnarray}
 {\cal H}_{\rm{eff}} &=& \frac{G_{F}}{\sqrt{2}}
     \bigg\{ \sum\limits_{q=u,c} V_{qb} V_{qD}^{*} \big[
     C_{1}({\mu}) O^{q}_{1}({\mu})
  +  C_{2}({\mu}) O^{q}_{2}({\mu})\Big]\nonumber\\
  &&-V_{tb} V_{tD}^{*} \Big[{\sum\limits_{i=3}^{10}} C_{i}({\mu}) O_{i}({\mu})
  \big ] \bigg\} + \mbox{H.c.} ,
 \label{eq:hamiltonian}
\end{eqnarray}
where $V_{qb(D)}$ and $V_{tb(D)}$ with $D=d,s$ are CKM matrix
elements. Functions $O_{i}$ ($i=1,...,10$) are local four-quark
operators :
 \begin{itemize}
 \item  current--current (tree) operators
    \begin{eqnarray}
  O^{q}_{1}=({\bar{q}}_{\alpha}b_{\beta} )_{V-A}
               ({\bar{D}}_{\beta} q_{\alpha})_{V-A},
    \ \ \ \ \ \ \ \ \
   O^{q}_{2}=({\bar{q}}_{\alpha}b_{\alpha})_{V-A}
               ({\bar{D}}_{\beta} q_{\beta} )_{V-A},
    \label{eq:operator12}
    \end{eqnarray}
     \item  QCD penguin operators
    \begin{eqnarray}
      O_{3}=({\bar{D}}_{\alpha}b_{\alpha})_{V-A}\sum\limits_{q^{\prime}}
           ({\bar{q}}^{\prime}_{\beta} q^{\prime}_{\beta} )_{V-A},
    \ \ \ \ \ \ \ \ \
    O_{4}=({\bar{D}}_{\beta} b_{\alpha})_{V-A}\sum\limits_{q^{\prime}}
           ({\bar{q}}^{\prime}_{\alpha}q^{\prime}_{\beta} )_{V-A},
    \label{eq:operator34} \\
     \!\!\!\! \!\!\!\! \!\!\!\! \!\!\!\! \!\!\!\! \!\!\!\!
    O_{5}=({\bar{D}}_{\alpha}b_{\alpha})_{V-A}\sum\limits_{q^{\prime}}
           ({\bar{q}}^{\prime}_{\beta} q^{\prime}_{\beta} )_{V+A},
    \ \ \ \ \ \ \ \ \
    O_{6}=({\bar{D}}_{\beta} b_{\alpha})_{V-A}\sum\limits_{q^{\prime}}
           ({\bar{q}}^{\prime}_{\alpha}q^{\prime}_{\beta} )_{V+A},
    \label{eq:operator56}
    \end{eqnarray}
 \item electro-weak penguin operators
    \begin{eqnarray}
     O_{7}=\frac{3}{2}({\bar{D}}_{\alpha}b_{\alpha})_{V-A}
           \sum\limits_{q^{\prime}}e_{q^{\prime}}
           ({\bar{q}}^{\prime}_{\beta} q^{\prime}_{\beta} )_{V+A},
    \ \ \ \
    O_{8}=\frac{3}{2}({\bar{D}}_{\beta} b_{\alpha})_{V-A}
           \sum\limits_{q^{\prime}}e_{q^{\prime}}
           ({\bar{q}}^{\prime}_{\alpha}q^{\prime}_{\beta} )_{V+A},
    \label{eq:operator78} \\
     O_{9}=\frac{3}{2}({\bar{D}}_{\alpha}b_{\alpha})_{V-A}
           \sum\limits_{q^{\prime}}e_{q^{\prime}}
           ({\bar{q}}^{\prime}_{\beta} q^{\prime}_{\beta} )_{V-A},
    \ \ \ \
    O_{10}=\frac{3}{2}({\bar{D}}_{\beta} b_{\alpha})_{V-A}
           \sum\limits_{q^{\prime}}e_{q^{\prime}}
           ({\bar{q}}^{\prime}_{\alpha}q^{\prime}_{\beta} )_{V-A},
    \label{eq:operator9x}
    \end{eqnarray}
\end{itemize}
where $\alpha$ and $\beta$ are color indices and $q^\prime$ are the
active quarks at the scale $m_b$, i.e. $q^\prime=(u,d,s,c,b)$. The
left-handed current is defined as $({\bar{q}}^{\prime}_{\alpha}
q^{\prime}_{\beta} )_{V-A}= {\bar{q}}^{\prime}_{\alpha} \gamma_\nu
(1-\gamma_5) q^{\prime}_{\beta}  $ and the right-handed current is
$({\bar{q}}^{\prime}_{\alpha} q^{\prime}_{\beta} )_{V+A}=
{\bar{q}}^{\prime}_{\alpha} \gamma_\nu (1+\gamma_5)
q^{\prime}_{\beta}$. The combinations $a_i$ of Wilson coefficients
are defined as usual \cite{Ali:1998eb}:
\begin{eqnarray}
a_1= C_2+C_1/3, &~a_2= C_1+C_2/3, &~ a_3= C_3+C_4/3,  ~a_4=
C_4+C_3/3,~a_5= C_5+C_6/3,\nonumber \\
a_6= C_6+C_5/3, &~a_7= C_7+C_8/3, &~a_8= C_8+C_7/3,~a_9=
C_9+C_{10}/3,
 ~a_{10}= C_{10}+C_{9}/3.
\end{eqnarray}

We work in the light-cone coordinate, in which a vector $V^{\mu}$ is
defined as $(\frac{V^0+V^3}{\sqrt{2}}, \frac{V^0-V^3}{\sqrt{2}},
V^1, V^2)$. We use $M_2$ to denote the charmed meson with a $c$
quark and $M_3$ to denote the meson with a $\bar c$ quark. In this
paper we work in the rest frame of $B$ meson and define the
direction in which $M_2$ moves as the positive direction of
$z$-axis. Therefore the momenta of $B_{(s)}$ meson and two charmed
mesons are defined in the light-cone coordinate as
\begin{eqnarray}
p_B=\frac{m_B}{\sqrt{2}}(1,1,\textbf{0}_{\perp}),\;
p_2=\frac{m_B}{\sqrt{2}}(1-r_3^2,r_2^2,\textbf{0}_{\perp}),\;
p_3=\frac{m_B}{\sqrt{2}}(r_3^2,1-r_2^2,\textbf{0}_{\perp}),\label{eq:momentum}
\end{eqnarray}
where $r_i=m_i/m_B$ ($i=2,3$) and $\textbf{0}_{\perp}$ are zero
two-component vectors. $m_2$ and $m_3$ are the masses of the two
charmed mesons. One can find that our definitions of the momentums
violate the on shell conditions. In the following calculations we
will keep the contributions of each diagram to the leading power of
$r_i$($i=2,3$). One will find that all the terms with a power of
$r_i$ higher than 2 are dropped. At this accuracy level, the
on-shell conditions can be satisfied. We use $k_1$, $k_2$, and $k_3$
to denote the momenta carried by the light quarks in $B_{(s)}$ meson
and two charmed mesons. They are defined by
\begin{eqnarray}
 k_1=(0,\frac{m_B}{\sqrt{2}}x_1,\textbf{k}_{1\perp}),\;
 k_2=(\frac{m_B}{\sqrt{2}}(1-r_3^2)x_2,0,\textbf{k}_{2\perp}),\;
 k_3=(0,\frac{m_B}{\sqrt{2}}(1-r_2^2)x_3,\textbf{k}_{3\perp}),\;\label{eq:fmoments}
\end{eqnarray}
with $x_1$, $x_2$ and $x_3$ as the momentum fractions.

\subsection{Wave functions of $B_{(s)}$ mesons}

The $B_{(s)}$ meson wave functions are
  decomposed into the following Lorentz structures:
 \begin{eqnarray}
 &&\int\frac{d^4z}{(2\pi)^4}e^{ik_1\cdot z}\langle0|\bar
 b_{\alpha}(0)d_{\beta}(z)|B_{(s)}(P_1)\rangle\nonumber\\
 &=&\frac{i}{\sqrt{2N_c}}\left\{(\not P_1+m_{B_{(s)}})\gamma_5[\phi_{B_{(s)}}(k_1)-\frac{\not n-\not v}{\sqrt{2}}
 \bar\phi_{B_{(s)}}(k_1)]\right\}_{\beta\alpha}.
 \end{eqnarray}
Here, $\phi_{B_{(s)}}(k_1)$ and $\bar\phi_{B_{(s)}}(k_1)$ are the
corresponding leading twist distribution amplitudes, and numerically
$\bar\phi_{B_{(s)}}(k_1)$ gives small contributions
~\cite{Lu:2002ny}, so we neglect it. The expression for
$\Phi_{B_{(s)}}$ becomes
 \begin{eqnarray}
 \Phi_{B_{(s)}}=\frac{i}{\sqrt{2N_c}}{(\not{P_1}+m_{B_{(s)}})\gamma_5\phi_{B_{(s)}}(k_1)}.
 \end{eqnarray}
 For the distribution amplitude in the b-space, we adopt the model
 function
 \begin{eqnarray}
  \phi_{B_{(s)}}(x,b)&=&N_{B_{(s)}}x^2(1-x)^2\exp\left[-\frac{1}{2}(\frac{xm_{B_{(s)}}}
  {\omega_b})^2-\frac{\omega_b^2b^2}{2}\right],\label{eq:Bwave}
  \end{eqnarray}
  where b is the conjugate space coordinate of
  $\textbf{k}_{1\perp}$. $N_{B_{(s)}}$ is the normalization constant,
  which is determined by the normalization condition
  \begin{eqnarray}
  \int^1_0 dx\phi_{B_{(s)}}(x,b=0)=\frac{f_{B_{(s)}}}{2\sqrt{2N_c}}.
  \end{eqnarray}
The $B^{\pm}$ and $B_d^0$ decays are studied intensively in PQCD
approach\cite{pqcd}. With the rich experimental data the
$\omega_b=0.40~\rm{GeV}$ is determined for $B$ meson. For $B_s$
meson, we will follow the authors in Ref.~\cite{bs} and adopt the
value $\omega_{b_s}=(0.50\pm0.05)~\rm{GeV}$.

\subsection{Wave function of $D^{(*)}/\bar D^{(*)}$ meson}
\label{subsec:Dwave}

In the heavy quark limit, the two-particle light-cone distribution
amplitudes of $D^{(*)}/\bar D^{(*)}$ meson are defined
as\cite{Kurimoto:2002sb}
 \begin{eqnarray}
\langle D(P_2)| q_{\alpha}(z) \bar c_{\beta}(0)|0\rangle
&=&\frac{i}{\sqrt{2N_C}}
 \int_0^1dxe^{ixP_2\cdot z}\left[\gamma_5(\not P_2 + m_D)
 \phi_D(x,b)\right]_{\alpha\beta}\nonumber\\
 \langle D^{*}(P_2)|
 q_{\alpha}(z)\bar c_{\beta}(0)|0\rangle&=&-\frac{1}{\sqrt{2N_C}}
 \int_0^1dxe^{ixP_2\cdot z}\left[\not{\epsilon}_L(\not{P_2}+m_{D^*})\phi_{D^*}^L(x,b)
+\not{\epsilon}_T(\not{P_2}+m_{D^*})\phi_{D^*}^T(x,b)\right]_{\alpha\beta}\label{definition_of_D}
 \end{eqnarray}
 with
 \begin{eqnarray}
  \int_0^1 dx\phi_D(x,0)&=&\frac{f_D}{2\sqrt{2N_c}}\;,
  \int_0^1dx\phi_{D^*}^L(x,0)=\frac{f_{D^*}}{2\sqrt{2N_c}}\;,
  \int_0^1dx\phi_{D^*}^T(x,0)=\frac{f_{D^*}^T}{2\sqrt{2N_c}}\;,
  \end{eqnarray}
  as the normalization conditions.
  In the heavy quark limit we
  have
  \begin{equation}
  f_{D^*}^T-f_{D^*}\frac{m_c+m_d}{M_{D^*}}\sim
  f_{D^*}-f_{D^*}^T\frac{m_c+m_d}{M_{D^*}}\sim
  O(\bar\Lambda/M_{D^*}).
  \end{equation}
 Thus we will use $f_{D^*}^T=f_{D^*}$ in our calculation.
 The model for the distribution amplitude for D
 meson that we used in this paper is
  \begin{eqnarray}
\phi_D(x,b) &=& \frac{1}{2\sqrt{2N_c}}f_D6x(1-x)[1+C_D(1-2x)]\exp
[\frac{-\omega^2 b^2}{2}],\label{eq:Dwave}
 \end{eqnarray}
 which has been tested in the $B\to D^{(*)}M$ and $B\to \bar
 D^{(*)}M$ decays \cite{Li:2008ts}.
 The masses of $D_{(s)}^{(*)}$ meson that we use are \cite{pdg}
 \begin{eqnarray}
m_{D}&=&1.869~\mbox{GeV}, \qquad m_{D_s^-} = 1.968~\mbox{GeV},  \nonumber \\
m_{D^{*}} &=& 2.010~\mbox{GeV}, \qquad m_{D_s^{*-}} =
2.112~\mbox{GeV}.
\end{eqnarray}
We use $C_D = 0.5\pm0.1$, $\omega = 0.1~\mbox{GeV}$ for $D/\bar{D}$
meson and $C_D = 0.4\pm0.1$, $\omega = 0.2~\mbox{GeV}$ for
$D_s/\bar{D}_s$ meson, which are determined in Ref.~\cite{Li:2008ts}
by fitting. In the wave function of $D^*_{(s)}$ mesons, the
$\phi_{D^*}^L$ and $\phi_{D^*}^T$ can not be related by the equation
of motion. We simply follow the authors in
Ref.~\cite{Kurimoto:2002sb} and adopt the same model as that of $D$
meson for them
\begin{equation}
\phi_{D^*}^L(x,b)=\phi_{D^*}^T(x,b)=\frac{1}{2\sqrt{2N_c}}f_{D^*}6x(1-x)[1+C_{D^*}(1-2x)]\exp
[\frac{-(\omega^*)^2 b^2}{2}].
\end{equation}
The mass difference of $D_{(s)}$ and $D_{(s)}^*$ is very small. In a
heavy quark limit, the light meson in $D_{(s)}^{(*)}$ mesons is not
sensitive to the spin and color of the heavy $c$ or $\bar c$ quark.
Thus the light-cone distribution amplitudes of $D_{(s)}$ and
$D_{(s)}^{*}$ should be very similar. In our calculation, we simply
take $C_{D^*}=C_D$ and $\omega^*=\omega$. $f_D =
(207\pm4)~\mbox{MeV}$ \cite{Follana:2007uv} and $f_{D_s} =
(241\pm3)~\mbox{MeV}$\cite{Follana:2007uv} are adopted and the
following relations derived from HQET \cite{Manohar:2000dt} are used
to determine $f_{D_{(s)}^*}$:
\begin{eqnarray}
f_{D^{*}} = \sqrt{\frac{m_{D}}{m_{D^*}}}f_{D}, \qquad f_{D_s^{*-}} =
\sqrt{\frac{m_{D_s^-}}{m_{D_s^*-}}}f_{D_s^-}.
\end{eqnarray}
The value of $f_{D_s}$ above is smaller than the recent experimental
data $f_{D_s}=(273\pm10)~\mbox{MeV}$~\cite{pdg}. Because the
amplitude in the PQCD approach is factorized as the convolution of
the wave functions, Sudakov factors and the hard part, the branching
ratio is proportional to the $f^2_{M_{2/3}}$. Thus if the
experimental data is adopted, our results for the branching ratios
will increase by $F=(\frac{273\pm10}{241\pm3})^2$ for single $D_s$
meson in the final state and $F^2$ for double $D_s$ meson final
state.

\subsection{Factorization Formulae for $B\to PP$ mode}
 \begin{figure}
 \begin{center}
 \includegraphics[width=7.cm]{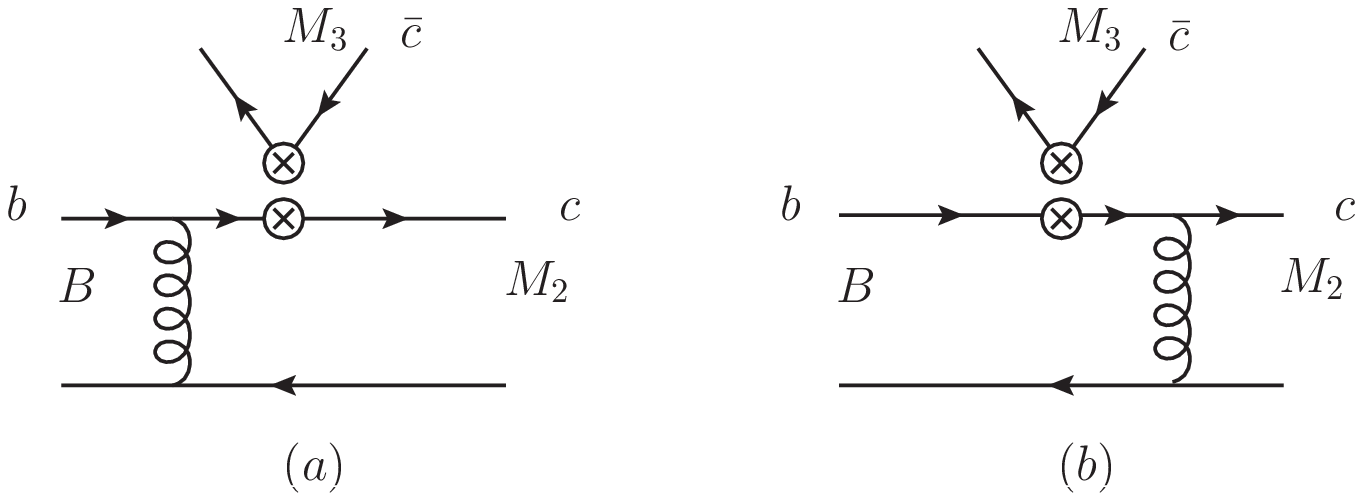}
 \hspace{1.0cm}
 \includegraphics[width=7.cm]{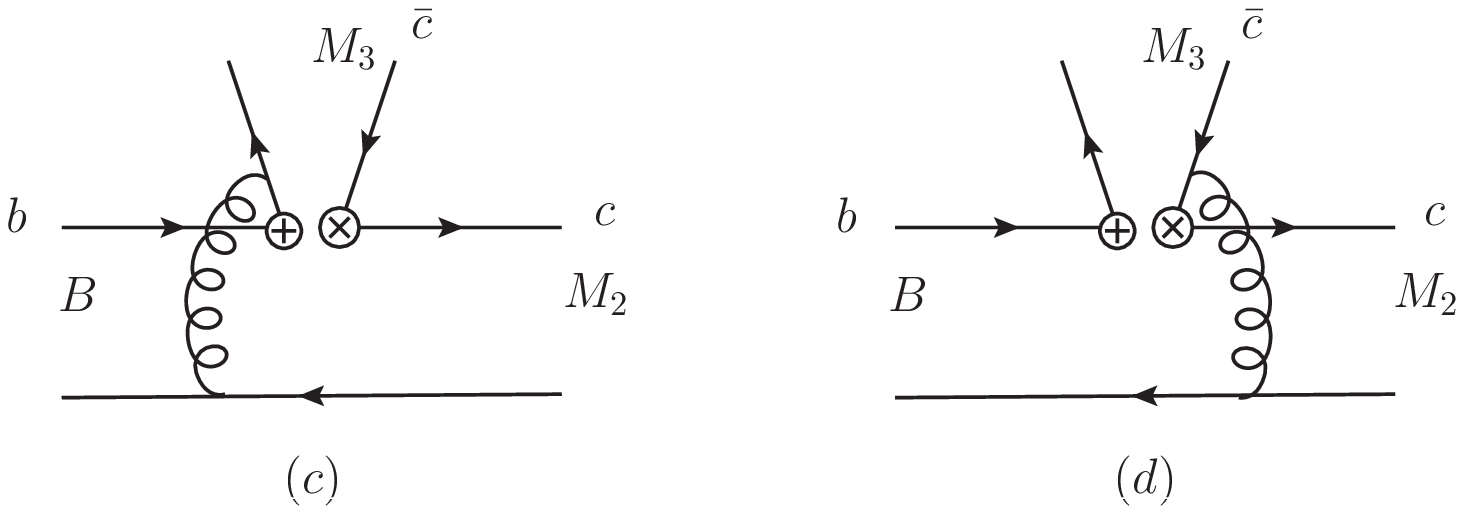}
\caption{Emission diagrams.}
 \label{fig:emission}
 \end{center}
 \end{figure}

\begin{figure}
 \begin{center}
 \includegraphics[width=5.cm]{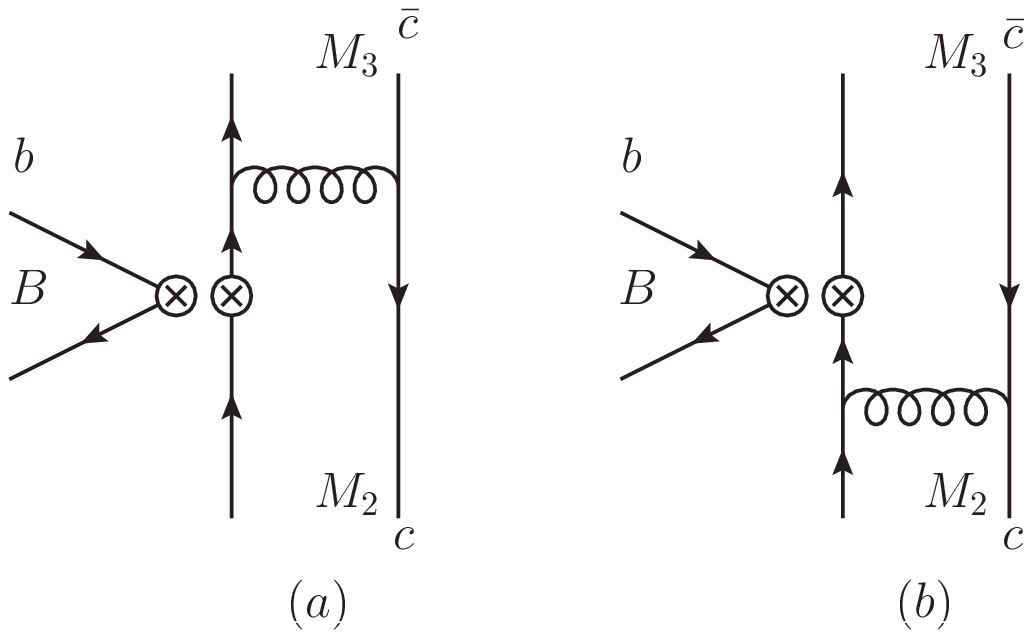}
 \hspace{1.0cm}
 \includegraphics[width=5.cm]{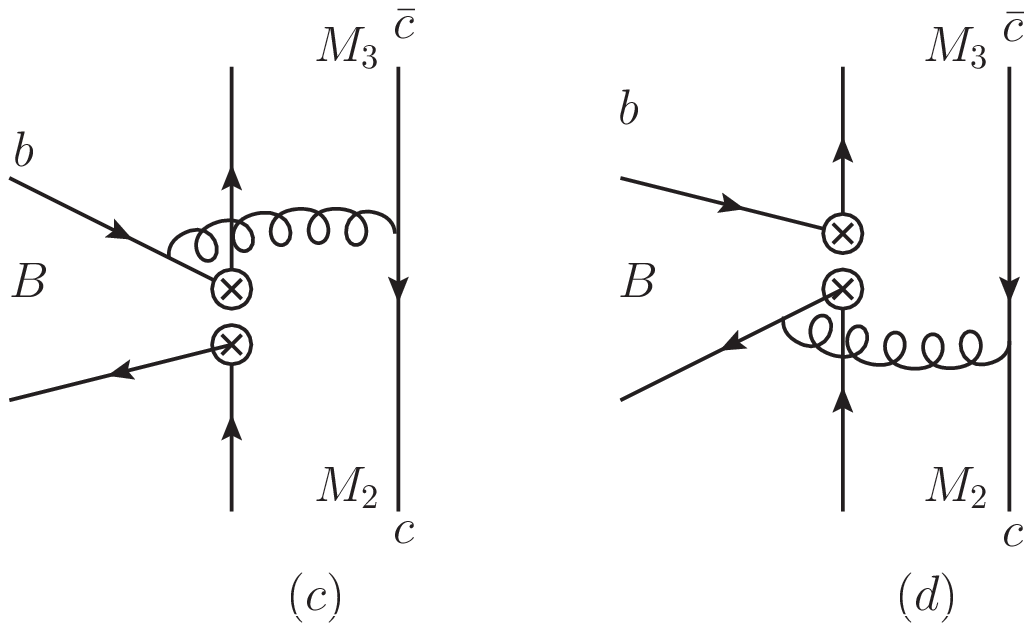}
\caption{Annihilation diagrams without charm quark in the four-quark
operator.}
 \label{fig:annihilation}
 \end{center}
 \end{figure}

\begin{figure}
 \begin{center}
 \includegraphics[width=5.cm]{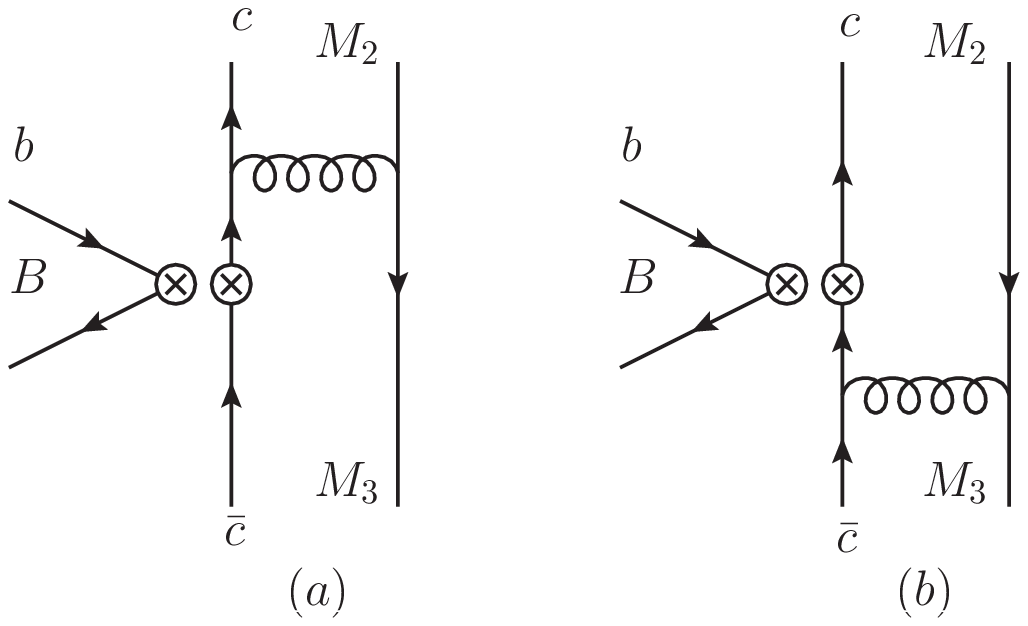}
 \hspace{1.0cm}
 \includegraphics[width=5.cm]{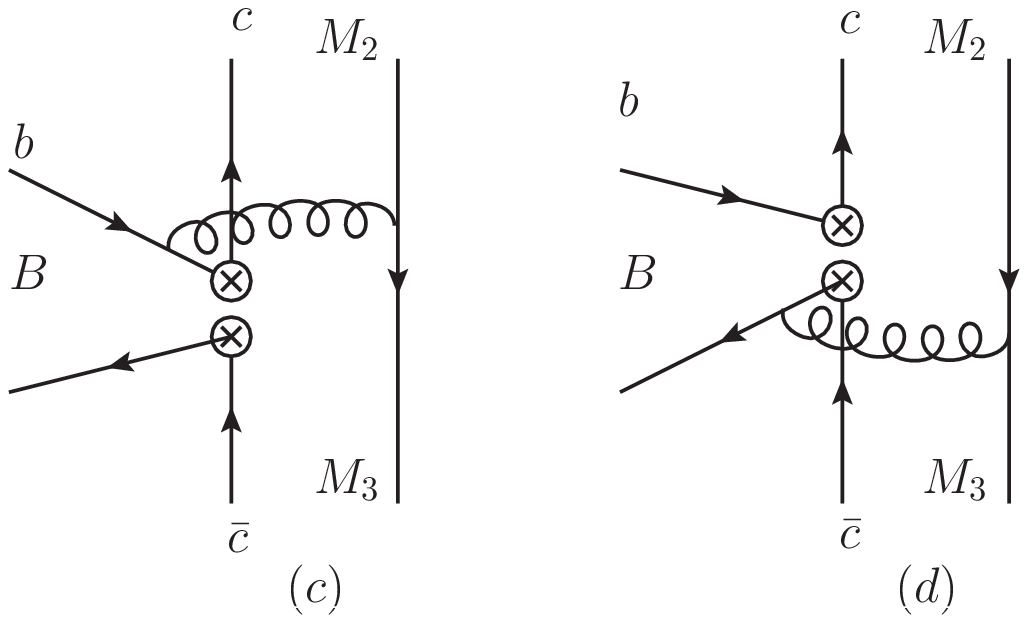}
\caption{Annihilation diagrams with charm quark in the four-quark
operator.}
 \label{fig:annihilationc}
 \end{center}
 \end{figure}

In this subsection, we list all the amplitudes from the Feynman
diagrams for $\langle M_2 M_3|C_i(\mu)O_i(\mu)|B_{(s)}\rangle$ up to
the leading order, with $M_2$ and $M_3$ as two charmed mesons.
According to their topological structures, the diagrams that
contribute to the decays of $B_{(s)}$ to two charmed mesons can be
divided into two types, the emission diagrams (see
Fig.~\ref{fig:emission}) with the light antiquark in $B_{(s)}$ meson
entering one of the charmed mesons as a spectator and the
annihilation diagrams (see Figs.~\ref{fig:annihilation} and
\ref{fig:annihilationc}) without any spectator quark. The first two
diagrams in Fig.~\ref{fig:emission} are factorizable diagrams, whose
amplitude can be naively factorized as a decay constant of a charmed
meson and a form factor like structure. The amplitudes arise from
all the possible Lorentz structure of the operators for factorizable
emission diagrams are given as following, where $a_i$ denotes the
Wilson coefficients and $t$ is the scale.
\begin{itemize}
\item Factorizable emission diagrams for (V-A)(V-A) operator
\begin{eqnarray}
 F_e^{LL}(a_i(t))&=&8\pi C_Ff_{M_3}m_B^4 \int_{0}^{1}d x_{1}d
 x_{2}\int_{0}^{1/\Lambda} b_1d b_1 b_2d b_2
 \phi_B(x_1,b_1)\phi_{M_2}(x_2)
 \nonumber \\
 &&\times \left[E_e(t_e^{(1)})a_i(t_e^{(1)})h_e(x_1,x_2(1-r_3^2),b_1,b_2)S_t(x_2)(1+x_2+r_2(1-2x_2))\right.\nonumber\\
 &&\left.+r_2(1+r_2)E_e(t_e^{(2)})a_i(t_e^{(2)})h_e(x_2,x_1(1-r_3^2),b_2,b_1)S_t(x_1)\right]\;.
\end{eqnarray}
\item Factorizable emission diagrams for (S-P)(S+P) operator
\begin{eqnarray}
 F_e^{SP}(a_i(t))&=&16\pi C_Ff_{M_3}m_B^4 \int_{0}^{1}d x_{1}d
 x_{2}\int_{0}^{1/\Lambda} b_1d b_1 b_2d b_2
 \phi_B(x_1,b_1)\phi_{M_2}(x_2)\nonumber \\
 & &\times r_3 \left[ E_e(t_e^{(1)})a_i(t_e^{(1)})h_e(x_1,x_2(1-r_3^2),b_1,b_2)S_t(x_2)(1+2r_2+r_2x_2)\right.\nonumber\\
 &&\left.+r_2E_e(t_e^{(2)})a_i(t_e^{(2)})h_e(x_2,x_1(1-r_3^2),b_2,b_1)S_t(x_1)\right]\;.
\end{eqnarray}
\end{itemize}
The amplitudes for the nonfactorizable emission diagrams in
Fig.\ref{fig:emission}(c) and (d) are given as:
\begin{itemize}
\item Nonfactorizable emission diagrams for (V-A)(V-A) operator are
\begin{eqnarray}
 F_{en}^{LL}(a_i(t))&=& 16\pi\sqrt{\frac{2}{3}} C_F m_B^4 \int_0^1
 [dx]\int_0^{1/\Lambda} b_1 db_1 b_3 db_3
 \phi_B(x_1,b_1)\phi_{M_2}(x_2)\phi_{M_3}(x_3)
 \nonumber \\
 & &\times
 \left[(x_3-r_2x_2)E_b(t_{en}^{(1)})a_i(t_{en}^{(1)})h^{(1)}_{en}(x_i,b_i)\right.\nonumber\\
 &&\left.+\big(x_2r_2-x_2+x_3-1\big)E_{en}(t_{en}^{(2)})a_i(t_{en}^{(2)})h^{(2)}_{en}(x_i,b_i)
 \right]\;.
\end{eqnarray}
\item Nonfactorizable emission diagrams for (V-A)(V+A) operator are
\begin{eqnarray}
 F_{en}^{LR}(a_i(t))&=& 16\pi\sqrt{\frac{2}{3}} C_F m_B^4 \int_0^1
 [dx]\int_0^{1/\Lambda} b_1 db_1 b_3 db_3
 \phi_B(x_1,b_1)\phi_{M_2}(x_2)\phi_{M_3}(x_3)
 \nonumber \\
 & &\times r_3 \left[(x_3+r_2(x_2+x_3))E_{en}(t_{en}^{(1)})a_i(t_{en}^{(1)})h^{(1)}_{en}(x_i,b_i)\right.\nonumber\\
 &&\left.-\big(r_2(x_2-x_3+2)-x_3+2\big)E_{en}(t_{en}^{(2)})a_i(t_{en}^{(2)})h^{(2)}_{en}(x_i,b_i)
 \right]\;.
\end{eqnarray}
\end{itemize}

The first two diagrams in Figs.~\ref{fig:annihilation} and
\ref{fig:annihilationc} are the factorizable diagrams for
annihilation diagrams, whose amplitudes can be factorized as a
$B_{(s)}$ meson decay constant and a form factor like structure
between two charmed mesons. It should be reminded that, in the
decays we considered, the factorizable annihilation diagrams can be
divided into two types, depending on whether the quark propagator is
a light quark propagator (see the first two diagrams in
Fig.~\ref{fig:annihilation}) or a $c$-quark propagator (see the
first diagrams in Fig.~\ref{fig:annihilationc}). In calculation we
keep the mass of the $c$-quark while the mass of the light quark is
neglected and thus these two types of diagrams have different
expressions. The amplitudes for the factorizable annihilation
diagrams with a light quark propagator (the first two diagrams in
Fig.~\ref{fig:annihilation}) are given as follows:
\begin{itemize}
\item Factorizable annihilation diagrams for (V-A)(V-A) operator
\begin{eqnarray}
 F_a^{LL}(a_i(t))&=&8\pi C_Ff_B m_B^4 \int_0^1
 dx_2dx_3\int_0^{1/\Lambda}b_2db_2b_3db_3 \phi_{M_2}(x_2) \phi_{M_3}(x_3)\nonumber \\
 & &\times \left[\big(2r_2r_3(x_2-2)+x_2-1\big)\right.\nonumber\\
 &&\left.\times E_a(t_a^{(1)})a_i(t_a^{(1)}) h_a(1-(1-r_2^2)x_3,1-(1-r_3^2)x_2,b_3,b_2)S_t(x_2)\right.\nonumber \\
 & &\left.+\big(-2r_2r_3(x_3-2)-(x_3-1)\big)\right.\nonumber\\
 &&\left.\times E_a(t_a^{(2)})a_i(t_a^{(2)})
 h_a(1-(1-r_3^2)x_2,1-(1-r_2^2)x_3,b_2,b_3)S_t(x_3)\right]\;.
\end{eqnarray}
The two terms of $F_a^{LL}(a_i(t))$ give destructive contributions.
Very little contribution appears when $\phi_{M_2}$ and $\phi_{M_3}$
are different from each other. Otherwise, $F_a^{LL}(a_i(t))=0$.
\item Factorizable annihilation diagrams for (V-A)(V+A) operator
\begin{eqnarray}
 F_a^{LR}(a_i(t))=F_a^{LL}(a_i(t)).
\end{eqnarray}
\item Factorizable annihilation diagrams for (S-P)(S+P) operator
\begin{eqnarray}
 F_a^{SP}(a_i(t))&=&16\pi C_Ff_B m_B^4 \int_0^1
 dx_2dx_3\int_0^{1/\Lambda}b_2db_2b_3db_3 \phi_{M_2}(x_2) \phi_{M_3}(x_3)\nonumber \\
 & &\times \left[(2r_3+r_2(1-x_2))E_a(t_a^{(1)})a_i(t_a^{(1)}) h_a(1-(1-r_2^2)x_3,1-(1-r_3^2)x_2,b_3,b_2)S_t(x_2)\right.\nonumber \\
 & &\left.+(2r_2+r_3(1-x_3))E_a(t_a^{(2)})a_i(t_a^{(2)})
 h_a(1-(1-r_3^2)x_2,1-(1-r_2^2)x_3,b_2,b_3)S_t(x_3)\right]\;.
\end{eqnarray}
\end{itemize}
For the amplitudes of the factorizable diagrams with a $c$-quark
propagator (the first two diagrams in Fig.~\ref{fig:annihilationc}),
we add the character ``c" in the subscript to distinguish them from
those with a light quark propagator. Because of current
conservation, the factorizable annihilation diagrams of $B\to PP$
decay mode for (V-A)(V-A) and (V-A)(V+A) operators cancel each
other. Amplitudes for these diagrams are given as follows:
\begin{itemize}
\item Factorizable annihilation diagrams for (V-A)(V-A) operator
\begin{eqnarray}
 F_{ac}^{LL}(a_i(t))&=&0\;.
\end{eqnarray}
\item Factorizable annihilation diagrams for (V-A)(V+A) operator
\begin{eqnarray}
 F_{ac}^{LR}(a_i(t))&=&F_{ac}^{LL}(a_i(t))=0.
\end{eqnarray}
\end{itemize}

Similar to the factorizable annihilation diagrams, the
nonfactorizable annihilation diagrams are also divided into two
types (see the last two diagrams of Figs.~\ref{fig:annihilation} and
\ref{fig:annihilationc}), depending on whether the $c\bar c$ are
generated from the effective weak vertex. Because the $c$ quark in
the charmed meson carries most of the energy, these two types of
nonfactorizable diagrams are expected to have different scales.
Additionally, because the momentum fraction $x_i$($i=2,3$) is
defined on the light quark in the charmed mesons, these two types of
nonfactorizable annihilation diagrams also have different
expressions. The amplitudes of the diagrams with $c\bar c$ pair
generated from a hard gluon (the last two diagrams in
Fig.~\ref{fig:annihilation}) are given as
\begin{itemize}
\item Nonfactorizable annihilation diagrams for (V-A)(V-A) operator
\begin{eqnarray}
 F_{an}^{LL}(a_i(t))&=& 16\pi\sqrt{\frac{2}{3}} C_F m_B^4 \int_0^1
 [dx]\int_0^{1/\Lambda}b_1 db_1 b_2db_2\phi_B(x_1,b_1)\phi_{M_2}(x_2)\phi_{M_3}(x_3) \nonumber \\
 & &\times \left[\big(r_2r_3(x_2+x_3-4)+x_3-1\big)
 E_{an}(t_{an}^{(1)})a_i(t_{an}^{(1)})h^{(1)}_{an}(x_i,b_i)\right.\nonumber\\
 &&\left.-\big(r_2r_3(x_2+x_3-2)+x_2-1\big) E_{an}(t_{an}^{(2)})a_i(t_{an}^{(2)})h^{(2)}_{an}(x_i,b_i)
 \right]\;.
\end{eqnarray}
\item Nonfactorizable annihilation diagrams for (V-A)(V+A) operator
\begin{eqnarray}
 F_{an}^{LR}(a_i(t))&=& 16\pi\sqrt{\frac{2}{3}} C_F m_B^4 \int_0^1
 [dx]\int_0^{1/\Lambda}b_1 db_1 b_2db_2\phi_B(x_1,b_1)\phi_{M_2}(x_2)\phi_{M_3}(x_3) \nonumber \\
 & &\times \left[-\big(r_2(x_2+1)-r_3(x_3+1)\big) E_{an}(t_{an}^{(1)})a_i(t_{an}^{(1)})h^{(1)}_{an}(x_i,b_i)\right.\nonumber\\
 &&\left.+ \big(r_2(x_2-1)-r_3(x_3-1)\big) E_{an}(t_{an}^{(2)})a_i(t_{an}^{(2)})h^{(2)}_{an}(x_i,b_i)
 \right]\;.
\end{eqnarray}
\item Nonfactorizable annihilation diagrams for (S-P)(S+P) operator
\begin{eqnarray}
 F_{an}^{SP}(a_i(t))&=& 16\pi\sqrt{\frac{2}{3}} C_F m_B^4 \int_0^1
 [dx]\int_0^{1/\Lambda}b_1 db_1 b_2db_2\phi_B(x_1,b_1)\phi_{M_2}(x_2)\phi_{M_3}(x_3) \nonumber \\
 & &\times \left[\big(r_2r_3(x_2+x_3-4)+x_2-1\big) E_{an}(t_{an}^{(1)})a_i(t_{an}^{(1)})h^{(1)}_{an}(x_i,b_i)\right.\nonumber\\
 &&\left. -\big(r_2r_3(x_2+x_3-2)+x_3-1\big) E_{an}(t_{an}^{(2)})a_i(t_{an}^{(2)})h^{(2)}_{an}(x_i,b_i)
 \right]\;.
\end{eqnarray}
\end{itemize}

Similar to what we do with the factorizable annihilation diagrams,
the amplitudes with $c\bar c$ pair from the effective weak vertex
are also distinguished by adding the character ``c" in the
subscripts. Amplitudes for these diagrams (the last two diagrams in
Fig.~\ref{fig:annihilationc}) are given by
\begin{itemize}
\item Nonfactorizable annihilation diagrams for (V-A)(V-A) operator
\begin{eqnarray}
 F_{anc}^{LL}(a_i(t))&=& 16\pi\sqrt{\frac{2}{3}} C_F m_B^4 \int_0^1
 [dx]\int_0^{1/\Lambda}b_1 db_1 b_2db_2\phi_B(x_1,b_1)\phi_{M_2}(x_2)\phi_{M_3}(x_3) \nonumber \\
 & &\times \left[\big(-r_2r_3(x_2+x_3+2)-x_2\big) E_{an}(t_{an}^{(1c)})a_i(t_{an}^{(1c)})h^{(1c)}_{an}(x_i,b_i)\right.\nonumber\\
 &&\left.+\big(r_2r_3(x_2+x_3)+x_3\big) E_{an}(t_{an}^{(2c)})a_i(t_{an}^{(2c)})h^{(2c)}_{an}(x_i,b_i)
 \right]\;.
\end{eqnarray}
\item Nonfactorizable annihilation diagrams for (S-P)(S+P) operator
\begin{eqnarray}
 F_{anc}^{SP}(a_i(t))&=& 16\pi\sqrt{\frac{2}{3}} C_F m_B^4 \int_0^1
 [dx]\int_0^{1/\Lambda}b_1 db_1 b_2db_2\phi_B(x_1,b_1)\phi_{M_2}(x_2)\phi_{M_3}(x_3) \nonumber \\
 & &\times \left[-\big(r_2r_3(x_2+x_3+2)+x_3\big) E_{an}(t_{an}^{(1c)})a_i(t_{an}^{(1c)})h^{(1c)}_{an}(x_i,b_i)\right.\nonumber\\
 &&\left.+\big((r_2r_3+1)x_2+r_3x_3(r_2+r_3)\big) E_{an}(t_{an}^{(2c)})a_i(t_{an}^{(2c)})h^{(2c)}_{an}(x_i,b_i)
 \right]\;.
\end{eqnarray}
\end{itemize}

\subsection{Analytic expressions for the decay amplitudes}
There are 10 decay channels for the $B\to PP$ decay mode, which can
be divided into two groups: decays with both emission and
annihilation contributions and pure annihilation type decays.
\begin{itemize}
\item Channels with both emission and annihilation contributions.
\begin{eqnarray}
{\cal A}(B^- \to D^0 D_{(s)}^-)&=&\frac{G_F}{\sqrt{2}}\left\{
V_{cb}V^*_{cd(s)}[F_e^{LL}(a_1)+F_{en}^{LL}(C_1)]+V_{ub}V^*_{ud(s)}[F_a^{LL}(a_1)+F_{an}^{LL}(C_1)]\right.\nonumber\\
&&\left.-V_{tb}V^*_{td(s)}[F_e^{LL}(a_4+a_{10})+F_{en}^{LL}(C_3+C_9)+F_e^{SP}(a_6+a_8)+F_{en}^{LR}(C_5+C_7)\right.\nonumber\\
&&\left.+F_a^{LL}(a_4+a_{10})+F_{an}^{LL}(C_3+C_9)+F_a^{SP}(a_6+a_8)+F_{an}^{LR}(C_5+C_7)]\right\},\label{eq:ampf}
\end{eqnarray}
\begin{eqnarray}
{\cal A}(\bar B^0 \to D^+ D^-)&=&\frac{G_F}{\sqrt{2}}\left\{
V_{cb}V^*_{cd}[F_e^{LL}(a_1)+F_{en}^{LL}(C_1)+F_{ac}^{LL}(a_2)+F_{anc}^{LL}(C_2)]\right.\nonumber\\
&&\left.-V_{tb}V^*_{td}[F_e^{LL}(a_4+a_{10})+F_{en}^{LL}(C_3+C_9)+F_e^{SP}(a_6+a_8)+F_{en}^{LR}(C_5+C_7)\right.\nonumber\\
&&\left.+F_{ac}^{LL}(a_3+a_9)+F_{anc}^{LL}(C_4+C_{10})+F_{ac}^{LR}(a_5+a_7)+F_{anc}^{SP}(C_6+C_8)\right.\nonumber\\
&&\left.+F_a^{LL}(a_3+a_4-\frac{1}{2}a_9-\frac{1}{2}a_{10})+F_{an}^{LL}(C_3+C_4-\frac{1}{2}C_9-\frac{1}{2}C_{10})\right.\nonumber\\
&&\left.+F_a^{LR}(a_5-\frac{1}{2}a_7)+F_{an}^{SP}(C_6-\frac{1}{2}C_8)
+F_a^{SP}(a_6-\frac{1}{2}a_8)+F_{an}^{LR}(C_5-\frac{1}{2}C_7)]\right\},
\end{eqnarray}
\begin{eqnarray}
{\cal A}(\bar B^0 \to D^+ D_s^-)&=&\frac{G_F}{\sqrt{2}}\left\{
V_{cb}V^*_{cs}[F_e^{LL}(a_1)+F_{en}^{LL}(C_1)]-V_{tb}V^*_{ts}[F_e^{LL}(a_4+a_{10})+F_{en}^{LL}(C_3+C_9)\right.\nonumber\\
&&\left.+F_e^{SP}(a_6+a_8)+F_{en}^{LR}(C_5+C_7)+F_a^{LL}(a_4-\frac{1}{2}a_{10})+F_{an}^{LL}(C_3-\frac{1}{2}C_9)\right.\nonumber\\
&&\left.+F_a^{SP}(a_6-\frac{1}{2}a_8)+F_{an}^{LR}(C_5-\frac{1}{2}C_7)]\right\},
\end{eqnarray}
\begin{eqnarray}
{\cal A}(\bar B_s^0 \to D_s^+ D^-)&=&\frac{G_F}{\sqrt{2}}\left\{
V_{cb}V^*_{cd}[F_e^{LL}(a_1)+F_{en}^{LL}(C_1)]-V_{tb}V^*_{td}[F_e^{LL}(a_4+a_{10})+F_{en}^{LL}(C_3+C_9)\right.\nonumber\\
&&\left.+F_e^{SP}(a_6+a_8)+F_{en}^{LR}(C_5+C_7)+F_a^{LL}(a_4-\frac{1}{2}a_{10})+F_{an}^{LL}(C_3-\frac{1}{2}C_9)\right.\nonumber\\
&&\left.+F_a^{SP}(a_6-\frac{1}{2}a_8)+F_{an}^{LR}(C_5-\frac{1}{2}C_7)]\right\},
\end{eqnarray}
\begin{eqnarray}
{\cal A}(\bar B_s^0 \to D_s^+ D_s^-)&=&\frac{G_F}{\sqrt{2}}\left\{
V_{cb}V^*_{cs}[F_e^{LL}(a_1)+F_{en}^{LL}(C_1)+F_{ac}^{LL}(a_2)+F_{anc}^{LL}(C_2)]\right.\nonumber\\
&&\left.-V_{tb}V^*_{ts}[F_e^{LL}(a_4+a_{10})+F_{en}^{LL}(C_3+C_9)+F_e^{SP}(a_6+a_8)+F_{en}^{LR}(C_5+C_7)\right.\nonumber\\
&&\left.+F_{ac}^{LL}(a_3+a_9)+F_{anc}^{LL}(C_4+C_{10})+F_{ac}^{LR}(a_5+a_7)+F_{anc}^{SP}(C_6+C_8)\right.\nonumber\\
&&\left.+F_a^{LL}(a_3+a_4-\frac{1}{2}a_9-\frac{1}{2}a_{10})+F_{an}^{LL}(C_3+C_4-\frac{1}{2}C_9-\frac{1}{2}C_{10})\right.\nonumber\\
&&\left.+F_a^{LR}(a_5-\frac{1}{2}a_7)+F_{an}^{SP}(C_6-\frac{1}{2}C_8)
+F_a^{SP}(a_6-\frac{1}{2}a_8)+F_{an}^{LR}(C_5-\frac{1}{2}C_7)]\right\},
\end{eqnarray}
\item Pure annihilation decays.
\begin{eqnarray}
{\cal A}(\bar B^0 \to D^0 \bar
D^0)&=&\frac{G_F}{\sqrt{2}}\left\{V_{cb}V^*_{cd}[F_{ac}^{LL}(a_2)+F_{anc}^{LL}(C_2)]
+V_{ub}V^*_{ud}[F_a^{LL}(a_2)+F_{an}^{LL}(C_2)]\right.\nonumber\\
&&\left.-V_{tb}V^*_{td}[F_a^{LL}(a_3+a_9)+F_{an}^{LL}(C_4+C_{10})+F_a^{LR}(a_5+a_7)+F_{an}^{SP}(C_6+C_{8})\right.\nonumber\\
&&\left.+F_{ac}^{LL}(a_3+a_9)+F_{anc}^{LL}(C_4+C_{10})+F_{ac}^{LR}(a_5+a_7)+F_{anc}^{SP}(C_6+C_{8})]\right\},
\end{eqnarray}
\begin{eqnarray}
{\cal A}(\bar B^0 \to D_s^+ D_s^-)&=&\frac{G_F}{\sqrt{2}}\left\{
V_{cb}V^*_{cd}[F_{ac}^{LL}(a_2)+F_{anc}^{LL}(C_2)]-V_{tb}V^*_{td}[F_{ac}^{LL}(a_3+a_9)+F_{anc}^{LL}(C_4+C_{10})\right.\nonumber\\
&&\left.+F_{ac}^{LR}(a_5+a_7)+F_{anc}^{SP}(C_6+C_8)+F_{a}^{LL}(a_3-\frac{1}{2}a_9)+F_{an}^{LL}(C_4-\frac{1}{2}C_{10})\right.\nonumber\\
&&\left.+F_{a}^{LR}(a_5-\frac{1}{2}a_7)+F_{an}^{SP}(C_6-\frac{1}{2}C_8)\right\},
\end{eqnarray}
\begin{eqnarray}
{\cal A}(\bar B_s^0 \to D^0\bar
D^0)&=&\frac{G_F}{\sqrt{2}}\left\{V_{cb}V^*_{cs}[F_{ac}^{LL}(a_2)+F_{anc}^{LL}(C_2)]
+V_{ub}V^*_{us}[F_a^{LL}(a_2)+F_{an}^{LL}(C_2)]\right.\nonumber\\
&&\left.-V_{tb}V^*_{ts}[F_a^{LL}(a_3+a_9)+F_{an}^{LL}(C_4+C_{10})+F_a^{LR}(a_5+a_7)+F_{an}^{SP}(C_6+C_{8})\right.\nonumber\\
&&\left.+F_{ac}^{LL}(a_3+a_9)+F_{anc}^{LL}(C_4+C_{10})+F_{ac}^{LR}(a_5+a_7)+F_{anc}^{SP}(C_6+C_{8})]\right\},
\end{eqnarray}
\begin{eqnarray}
{\cal A}(\bar B_s^0 \to D^+ D^-)&=&\frac{G_F}{\sqrt{2}}\left\{
V_{cb}V^*_{cs}[F_{ac}^{LL}(a_2)+F_{anc}^{LL}(C_2)]-V_{tb}V^*_{ts}[F_{ac}^{LL}(a_3+a_9)+F_{anc}^{LL}(C_4+C_{10})\right.\nonumber\\
&&\left.+F_{ac}^{LR}(a_5+a_7)+F_{anc}^{SP}(C_6+C_8)+F_{a}^{LL}(a_3-\frac{1}{2}a_9)+F_{an}^{LL}(C_4-\frac{1}{2}C_{10})\right.\nonumber\\
&&\left.+F_{a}^{LR}(a_5-\frac{1}{2}a_7)+F_{an}^{SP}(C_6-\frac{1}{2}C_8)\right\}.\label{eq:ampl}
\end{eqnarray}
\end{itemize}

There are also 10 decay channels for each category of $B\to PV$,
$B\to VP$, and $B\to VV$ decays. The decay amplitudes of the $B\to
PV$ and $B\to VP$ modes can be obtained from the $B\to PP$ decays
just by substituting the $D_{(s)}/\bar D_{(s)}$ meson for the
corresponding $D^*_{(s)}/\bar D^*_{(s)}$ meson. The factorization
formulae for these two decay modes are listed in
Appendix~\ref{appendix:BtoPV} and~\ref{appendix:BtoVP},
respectively.

The amplitude of $B\to VV$ decay can be decomposed as
\begin{eqnarray}
 {\cal A}(\epsilon_{2},\epsilon_{3})&=&i{\cal A}^N + i(\epsilon^*_{2T} \cdot \epsilon^*_{3T}){\cal A}^s
 + (\epsilon_{\mu \nu \alpha \beta}n^{\mu} \bar n^{\nu} \epsilon^{*\alpha}_{2T} \epsilon^{*\beta}_{3T}) {\cal A}^p,
\end{eqnarray}
where ${\cal A}^N$, including the D wave and part of the S wave
component, contains the contribution from the longitudinal
polarizations ${\cal A}^s$ and ${\cal A}^p$, corresponding to part
of the S wave component and all the P wave component, respectively,
which represent the transversely polarized contributions, and they
have the following relationships with the helicity amplitudes (an
$i$ in the amplitude is dropped):
 \begin{eqnarray}
 A_0={\cal A}^N\;,\; A_{\pm}={\cal A}^s \pm {\cal A}^p\;.
 \end{eqnarray}
For each decay process of $B\to VV$, the amplitudes ${\cal A}^N$,
${\cal A}^s$, and ${\cal A}^p$ have the same structures as
Eq.(\ref{eq:ampf})-(\ref{eq:ampl}), respectively. The factorization
formulae for the longitudinal and transverse polarization for the
$B\to VV$ decays are all listed in Appendix~\ref{appendix:BtoVV}.

\section{Numerical Analysis}
\label{sec:numerical}
The decay widths of $B$ to two charmed mesons decays can be directly
derived from the formulas of two-body decays in Ref.~\cite{pdg}.
With the amplitude obtained in Sec.~\ref{sec:analytic}, the decay
widths for the $B\to PP$, $B\to PV$, and $B\to VP$ decays are given
by
\begin{eqnarray}
\Gamma=\frac{[(1-(r_2+r_3)^2)(1-(r_2-r_3)^2)]^{1/2}}{16\pi
m_B}|{\cal A}|^2.
\end{eqnarray}
For the $B\to VV$ decays, the decay width is given by
\begin{eqnarray}
\Gamma&=&\frac{[(1-(r_2+r_3)^2)(1-(r_2-r_3)^2)]^{1/2}}{16\pi
m_B}\sum_{i=0,+,-}|A_i|^2.
\end{eqnarray}
The branching ratio is given by ${\cal BR}=\Gamma\tau_B$.

The key observables of the decays related in this paper are the CP
averaged branching ratios as well as direct CP
asymmetries($A_{\rm{CP}}^{\rm{dir}}$) and mixing induced CP
asymmetries($A_{\rm{CP}}^{\rm{mix}}$). Readers are referred to
Ref.~\cite{CPV} for some reviews on CP violation. First, we define
four amplitudes as follows:
\begin{eqnarray}
 A_f&=&\langle f|{\cal H}|B\rangle,\;\;\;
 \bar A_f=\langle f|{\cal H}|\bar B\rangle,\nonumber\\
 A_{\bar f}&=&\langle\bar f|{\cal H}|B\rangle,\;\;\;
 \bar A_{\bar f}=\langle\bar f|{\cal H}|\bar B\rangle,
\end{eqnarray}
where $\bar B$ meson has a $b$ quark in it and $\bar f$ is the CP
conjugate state of $f$. The direct CP asymmetry
$A_{\rm{CP}}^{\rm{dir}}$ is defined by
\begin{eqnarray}
 A_{\rm{CP}}^{\rm{dir}}&=&\frac{|\bar A_{\bar f}|^2-|A_f|^2}
                         {|\bar A_{\bar f}|^2+|A_f|^2}\;.
\end{eqnarray}
In neutral $B$ meson decays, if the final states are CP eigen states
$f =\bar f$, the time-dependent CP asymmetry   with mixing effects
present, is defined by
\begin{eqnarray}
 A_{\rm{CP}}(B(t)\to f)&\equiv&\frac{\Gamma(B(t)\to f)-\Gamma(\bar
 B(t)\to f)}{\Gamma(B(t)\to f)+\Gamma(\bar B(t)\to f)}\nonumber\\
 &=&-C_f\cos(\Delta Mt)
 +A_{\rm{CP}}^{\rm{mix}}(B\to f)\sin(\Delta Mt),
\end{eqnarray}
where $\Delta M$ is the mass difference of $B$ meson mass
eigenstates. After some calculation, we can get the explicit
expressions
\begin{eqnarray}
 C_f&=&{|A_f|^2-|\bar A_f|^2 \over |A_f|^2+|\bar A_f|^2}\;,\nonumber\\
 A_{\rm{CP}}^{\rm{mix}}&=&\frac{2Im[\frac{q}{p}\bar A_f A^*_f]}{|A_f|^2+|\bar
 A_f|^2}\;.
\end{eqnarray}
Since the mixing CP violation in neutral B meson system is
negligible in a good approximation, we have
\begin{eqnarray}
 \frac{q}{p}=e^{-i\phi_{M(B)}}={V^*_{tb}V_{t(d/s)} \over
 V_{tb}V^*_{t(d/s)}}.\label{2beta}
\end{eqnarray}
  Our
results for CP averaged branching ratios and CP asymmetries are
listed in Tables \ref{tab:BRforBtoPP}, \ref{tab:BRforBtoPV},
\ref{tab:BRforBtoVP}, \ref{tab:BRforBtoVV}, and
\ref{tab:CPforBtoVV}. All the experimental data are from the
Particle Data Group\cite{pdg} except the ones marked with ``BaBar''
and ``Belle''. In Table \ref{tab:BRforBtoVV}, we also list the
ratios of the transverse polarizations ${\cal R}_T$ in the branching
ratios for $B\to VV$ decays, which is defined by
\begin{eqnarray}
 {\cal R}_T&=&\frac{|A_+|^2+|A_-|^2}{|A_0|^2+|A_+|^2+|A_-|^2}\;.
\end{eqnarray}
The first errors in our results are estimated from the hadronic
parameters: (1) The decay constants of $B_{(s)}$ mesons:
$f_B=(0.19\pm 0.025)\rm{GeV}$ for $B$ mesons and $f_{B_s}=(0.23\pm
0.03)\rm{GeV}$ for $B_s$ meson; (2) The shape parameters in
$B_{(s)}$ meson wave functions: $\omega_b=(0.40\pm0.05)\rm{GeV}$ for
$B$ meson and $\omega_{b_s}=(0.50\pm0.05)\rm{GeV}$ for $B_s$ meson;
(3) The decay constants and the shape parameters in the wave
functions of charmed mesons, which are given in the last paragraph
in Sec.~\ref{subsec:Dwave}. The second errors are from the not known
next-to-leading order QCD corrections with respect to $\alpha_s$ and
nonperturbative power corrections with respect to scales in Eq.(7),
characterized by
 the choice of
the $\Lambda_{\rm{QCD}}=(0.25\pm0.05)\rm{GeV}$ and the variations of
the factorization scales  shown in Appendix~\ref{appendix:scale&h}.
The third errors are brought in by the CKM matrix elements, which
are given as\cite{CKMfitter}
\begin{eqnarray}
 |V_{cb}|&=&0.041\;17_{-0.001\;15}^{+0.000\;38}\;,\;|V_{cd}|=0.225\;08_{-0.000\;82}^{+0.000\;82}\;,\;|V_{cs}|=0.973\;47_{-0.000\;19}^{+0.000\;19}\;,\;\nonumber\\
 |V_{ub}|&=&0.003\;5_{-0.000\;14}^{+0.000\;15}\;,\;|V_{ud}|=0.974\;44_{-0.000\;28}^{+0.000\;28}\;\;,|V_{us}|=0.225\;7_{-0.001\;1}^{+0.001\;1}\;,\;\nonumber\\
 |V_{tb}|&=&0.999\;146_{-0.000\;016}^{+0.000\;047}\;,\;|V_{td}|=0.008\;59_{-0.000\;29}^{+0.000\;27}\;,\;|V_{ts}|=0.040\;41_{-0.001\;15}^{+0.000\;38}\;,\;\nonumber\\
 \gamma&=&(67.8_{-3.9}^{+4.2})^\circ\;,\;\beta=(21.58_{-0.81}^{+0.91})^\circ\;.
 \end{eqnarray}
 The other input parameters are \cite{pdg}
 \begin{eqnarray}
 G_F&=&1.16639\times10^{-5}\rm{GeV}^{-2}\;,\;\nonumber\\
 \tau_{B^-}&=&1.639\times10^{-12}s/\hbar\;,\;\tau_{B^0}=1.530\times10^{-12}s/\hbar\;,\;\tau_{B_s^0}=1.478\times10^{-12}s/\hbar\;,\;\nonumber\\
 m_B&=&5.28\rm{GeV}\;,\;m_{B_s}=5.366\rm{GeV}\;,\;m_D=1.87\rm{GeV}\;,\;m_{D_s}=1.97\rm{GeV}\;,\;\nonumber\\
 m_{D^*}&=&2.01\rm{GeV}\;,\;m_{D^*_s}=2.11\rm{GeV}\;,\;\hbar=6.582119\times10^{-25}\mbox{GeV s}\;.
\end{eqnarray}
Because in the direct CP asymmetries the errors arising from the CKM
elements are very small, we neglect them. In the $B\to VV$ decays,
the ratios of the transverse polarizations' contributions (${\cal
R}_T$)are not very sensitive to the parameters listed above. The
next-to-leading order corrections on r occur at the $r^2=0.13$ order
and thus the errors from the higher orders of r are very small
except for ${\cal R}_T$. This is confirmed at the numerical
calculations. Therefore we only keep these errors in ${\cal R}_T$
and neglect them in other physical quantities. We will talk about
the errors of these ratios later.

 \begin{table}
 \caption{CP averaging branching ratios (unit: $10^{-4}$) and the CP asymmetries  for $B\to PP$ decays.}
 \label{tab:BRforBtoPP}
 \begin{center}
 \begin{tabular}{cc|cc|cc|cc}
 \hline\hline
 &   &\multicolumn{2}{c|}{$\cal BR$}     &\multicolumn{2}{c|}{$A_{CP}^{\rm{dir}}(\%)$}      &\multicolumn{2}{c}{$A_{\rm{CP}}^{\rm{mix}}$}   \\
 \hline
 \multicolumn{2}{c|}{Channels}              &{Exp.}     &{This work}      &{Exp.}     &{This work}    &{Exp.}     &{This work}\\
 \hline
\ \ \ 1 &$B^- \to D^0 D^-$                &$4.2\pm0.6$   &$3.9_{-1.9-1.1-0.2}^{+2.9+0.7+0.1}$    &$-13\pm14\pm2$      &$0.6_{-0.0-0.1}^{+0.4+0.4}$   &$$             &$...$\\
\ \ \ 2 &$B^- \to D^0 D_{s}^-$            &$103\pm17$    &$95_{-46-26-6}^{+69+18+2}$   &$$                    &$\sim -10^{-3}$  &$$         &$...$\\
\ \ \ 3 &$\bar B^0 \to D^+ D^-$           &$2.11\pm0.31$ &$3.7_{-1.8-0.9-0.2}^{+2.9+0.4+0.1}$    &$11\pm22\pm7$[Babar]&$0.5_{-0.2-0.4}^{+0.1+0.5}$ &$-0.81\pm0.29$           &$-0.73_{-0.00-0.01-0.02}^{+0.00+0.01+0.02}$\\
\ \ \   &$$           &$$                                     &   &$-91\pm23\pm6$[Belle]  &$$       &$$      \\
\ \ \ 4 &$\bar B^0 \to D^+ D_s^-$         &$74\pm7$      &$89_{-43-25-5}^{+68+18+2}$     &$$&$...$ &$$                 &$...$\\
\ \ \ 5 &$\bar B_s^0 \to D_s^+ D^-$       &$$            &$2.2_{-1.0-0.7-0.1}^{+1.4+0.7+0.1}$       &$$               &$0.5_{-0.0-0.1}^{+0.1+0.2}$      &$$                 &$...$\\
\ \ \ 6 &$\bar B_s^0 \to D_s^+ D_s^-$     &$110\pm40$    &$55_{-24-15-3}^{+36+12+1}$          &$$&$...$ &$$                  &$...$\\
 \hline
\ \ \ 7 &$\bar B^0 \to D^0 \bar D^0$      &$<0.6$[BaBar]    &$0.28_{-0.11-0.08-0.02}^{+0.07+0.03+0.01}$        &$$  &$-5.3_{-2.7-3.3-0.3}^{+0.2+0.0+0.2}$    &$$            &$-0.74_{-0.01-0.01-0.02}^{+0.00+0.00+0.02}$\\
\ \ \ 8 &$\bar B^0 \to D_s^+ D_s^-$       &$< 0.36$[Belle]  &$0.35_{-0.13-0.10-0.02}^{+0.12+0.07+0.01}$      &$$    &$-2.3_{-0.4-0.4}^{+0.5+0.8}$    &$$           &$-0.73_{-0.00-0.01-0.02}^{+0.00+0.00+0.02}$\\
\ \ \ 9 &$\bar B_s^0 \to D^0\bar D^0$     &$$ &$5.0_{-1.5-1.2-0.3}^{+1.7+1.0+0.1}$ &$$             &$0.2_{-0.0-0.0}^{+0.1+0.1}$   &$$         &$\sim 10^{-3}$\\
\ \ \ 10 &$\bar B_s^0 \to D^+ D^-$         &$$ &$5.2_{-1.9-1.4-0.3}^{+1.5+0.7+0.1}$ &$$              &$...$   &$$          &$...$\\
 \hline\hline
 \end{tabular}
 \end{center}
 \end{table}

 \begin{table}
 \caption{CP averaging branching ratios (unit: $10^{-4}$) and the CP asymmetries for $B\to PV$ decays.}
 \label{tab:BRforBtoPV}
 \begin{center}
 \begin{tabular}{cc|cc|cc}
 \hline\hline
       &       &\multicolumn{2}{c|}{${\cal BR}$}       &\multicolumn{2}{c}{$A_{CP}^{\rm{dir}}(\%)$}        \\
 \hline
 \multicolumn{2}{c|}{Channels}               &{Exp.}    &{This work}     &{Exp.}    &{This work}   \\
 \hline
 1 &$B^- \to D^0 D^{*-}$                &$3.9\pm0.5$  &$3.6_{-1.7-1.0-0.2}^{+2.6+0.7+0.1}$ &$$         &$0.1_{-0.1-0.1}^{+0.4+0.1}$      \\
 2 &$B^- \to D^0 D_{s}^{*-}$            &$78\pm16$    &$89_{-42-24-5}^{+64+20+2}$  &$$       &$\sim -10^{-3}$           \\
 3 &$\bar B^0 \to D^+ D^{*-}$           &$6.1\pm1.5$  &$3.2_{-1.5-0.8-0.2}^{+2.4+0.5+0.1}$ &$-6\pm9$    &$\sim 10^{-2}$  \\
 4 &$\bar B^0 \to D^+ D_s^{*-}$         &$76\pm16$    &$83_{-39-23-5}^{+61+17+2}$  &$$        &$...$        \\
 5 &$\bar B_s^0 \to D_s^+ D^{*-}$       &$$           &$2.1_{-0.9-0.7-0.1}^{+1.3+0.7+0.0}$ &$$                       &$0.1_{-0.1-0.0}^{+0.0+0.0}$       \\
 6 &$\bar B_s^0 \to D_s^+ D_s^{*-}$     &$$           &$48_{-21-15-3}^{+31+15+1}$  &$$                        &$...$        \\
 \hline
 7 &$\bar B^0 \to D^0 \bar D^{*0}$      &$<2.9$[Babar]\cite{Aubert:2006ia}  &$(4.6_{-1.7-1.4-0.2}^{+1.5+1.3+0.9})\times 10^{-2}$   &$$         &$-4.1_{-4.4-2.9}^{+1.3+0.0}$ \\
 8 &$\bar B^0 \to D_s^+ D_s^{*-}$       &$<1.3$[BaBar]\cite{Aubert:2005jv}  &$(3.5_{-1.2-1.1-0.2}^{+1.4+1.8+0.1})\times 10^{-2}$  &$$          &$0.5_{-0.3-0.7}^{+0.1+1.7}$  \\
 9 &$\bar B_s^0 \to D^0\bar D^{*0}$     &$$                                 &$0.83_{-0.24-0.19-0.04}^{+0.41+0.32+0.01}$   &$$         &$0.4_{-0.2-0.1}^{+0.1+0.1}$  \\
 10 &$\bar B_s^0 \to D^+ D^{*-}$         &$$                                &$0.74_{-0.29-0.23-0.04}^{+0.23+0.24+0.01}$  &$$          &$...$ \\
 \hline\hline
 \end{tabular}
 \end{center}
 \end{table}

 \begin{table}
 \caption{CP averaging branching ratios  (unit: $10^{-4}$) and the CP asymmetries for $B\to VP$ (characterized by $B\to V$ form factor) decays.}
 \label{tab:BRforBtoVP}
 \begin{center}
 \begin{tabular}{cc|cc|cc}
 \hline\hline
        &       &\multicolumn{2}{c|}{${\cal BR}$}       &\multicolumn{2}{c|}{$A_{CP}^{\rm{dir}}(\%)$}        \\
 \hline
 \multicolumn{2}{c|}{Channels}               &{Exp.}    &{This work}     &{Exp.}    &{This work}   \\
 \hline
 1 &$B^- \to D^{*0} D^-$                &$6.3\pm1.4\pm1.0$[BaBar]\cite{Aubert:2006ia} &$4.8_{-2.3-1.4-0.3}^{+3.4+1.1+0.1}$ &$$             &$-0.5_{-0.2-0.3}^{+0.1+0.0}$  \\
 2 &$B^- \to D^{*0} D_{s}^-$            &$84\pm17$                                    &$119_{-56-34-7}^{+94+27+2}$ &$$           &$\sim -10^{-3}$   \\
 3 &$\bar B^0 \to D^{*+} D^-$           &$8.8\pm1.6$                                  &$4.6_{-2.1-1.1-0.3}^{+3.5+0.9+0.1}$     &$$           &$-0.6_{-0.1-0.2}^{+0.0+0.1}$ \\
 4 &$\bar B^0 \to D^{*+} D_s^-$         &$83\pm11$                                    &$112_{-53-32-6}^{+86+26+2}$   &$$            &$...$      \\
 5 &$\bar B_s^0 \to D_s^{*+} D^-$       &$$                                           &$2.7_{-1.1-0.9-0.1}^{+1.7+0.9+0.1}$    &$$            &$-0.4_{-0.0-0.1}^{+0.0+0.1}$     \\
 6 &$\bar B_s^0 \to D_s^{*+} D_s^-$     &$$                                           &$70_{-31-21-4}^{+44+19+1}$     &$$           &$...$       \\
 \hline
 7 &$\bar B^0 \to D^{*0} \bar D^0$      &$$                                  &$0.21_{-0.08-0.05-0.01}^{+0.06+0.03+0.00}$  &$$         &$-1.2_{-1.2-1.4-0.1}^{+0.3+0.7+0.1}$     \\
 8 &$\bar B^0 \to D_s^{*+} D_s^-$       &$<1.3$[BaBar]\cite{Aubert:2005jv}   &$0.25_{-0.08-0.08-0.01}^{+0.08+0.06+0.01}$   &$$        &$\sim -10^{-3}$    \\
 9 &$\bar B_s^0 \to D^{*0}\bar D^0$     &$$                                  &$4.3_{-1.3-1.1-0.2}^{+1.3+0.8+0.1}$  &$$          &$0.2_{-0.1-0.1}^{+0.0+0.0}$    \\
 10 &$\bar B_s^0 \to D^{*+} D^-$         &$$                                 &$4.4_{-1.3-1.2-0.2}^{+1.4+0.9+0.1}$  &$$          &$...$   \\
 \hline\hline
 \end{tabular}
 \end{center}
 \end{table}

 \begin{table}
 \caption{CP averaging branching ratios for $B\to VV$(unit: $10^{-4}$) and
 the ratios of the transverse polarizations' contribution.}
 \label{tab:BRforBtoVV}
 \begin{center}
  \begin{tabular}{cc|cc|cc}
 \hline\hline
       &       &\multicolumn{2}{c}{${\cal BR}$}       &\multicolumn{2}{c}{${\cal R}_T$}      \\
 \hline
 \multicolumn{2}{c}{Channels}               &{Exp.}  &{This work}    &{Exp.}  &{This work}    \\
 \hline
 1 &$B^- \to D^{*0} D^{*-}$                &$8.1\pm1.2\pm1.2$[Babar]\cite{Aubert:2006ia}  &$6.8_{-3.2-2.0-0.4}^{+5.0+1.5+0.1}$  &$$                    &$0.45_{-0.13}^{+0.13}$ \\
 2 &$B^- \to D^{*0} D_{s}^{*-}$            &$175\pm23$                                    &$181_{-95-53-10}^{+139+41+3.5}$  &$$                        &$0.48_{-0.14}^{+0.12}$ \\
 3 &$\bar B^0 \to D^{*+} D^{*-}$           &$8.2\pm0.9$                                   &$6.3_{-3.0-1.6-0.4}^{+4.8+1.1+0.1}$  &$0.43\pm0.08\pm0.02$  &$0.46_{-0.14}^{+0.14}$ \\
 4 &$\bar B^0 \to D^{*+} D_s^{*-}$         &$179\pm14$                                    &$168_{-88-48-9.6}^{+130+39+3.2}$  &$0.48\pm0.05$            &$0.48_{-0.14}^{+0.13}$ \\
 5 &$\bar B_s^0 \to D_s^{*+} D^{*-}$       &$$                                            &$3.9_{-1.9-1.3-0.2}^{+2.6+1.2+0.1}$  &$$                    &$0.44_{-0.14}^{+0.14}$ \\
 6 &$\bar B_s^0 \to D_s^{*+} D_s^{*-}$     &$$                                            &$99_{-54-29-5.6}^{+72+26+1.9}$   &$$                        &$0.47_{-0.15}^{+0.15}$ \\
 \hline
 7 &$\bar B^0 \to D^{*0} \bar D^{*0}$      &$<0.9$[Babar]\cite{Aubert:2006ia}       &$0.15_{-0.04-0.03-0.01}^{+0.05+0.03+0.00}$        &$$             &$0.47_{-0.29}^{+0.35}$   \\
 8 &$\bar B^0 \to D_s^{*+} D_s^{*-}$       &$<2.4$[Babar]\cite{Aubert:2005jv}       &$0.19_{-0.07-0.05-0.01}^{+0.10+0.06+0.00}$       &$$              &$0.57_{-0.37}^{+0.33}$ \\
 9 &$\bar B_s^0 \to D^{*0}\bar D^{*0}$     &$$                                      &$2.8_{-0.8-0.6-0.2}^{+1.1+0.7+0.1}$       &$$                     &$0.48_{-0.27}^{+0.37}$  \\
 10 &$\bar B_s^0 \to D^{*+} D^{*-}$         &$$                                     &$3.1_{-0.8-0.7-0.2}^{+1.0+0.9+0.1}$       &$$                     &$0.49_{-0.29}^{+0.34}$  \\
 \hline\hline
 \end{tabular}
 \end{center}
 \end{table}

 \begin{table}
 \caption{CP asymmetry and the ratios of P-wave contributions in branching ratios for $B\to VV$ decays.}
 \label{tab:CPforBtoVV}
 \begin{center}
  \begin{tabular}{cc|cc|cc|c}
 \hline\hline
 $$      &$$         &\multicolumn{2}{c|}{$A_{CP}^{\rm{dir}}(\%)$}   &\multicolumn{2}{c}{$A_{\rm{CP}}^{\rm{mix}}$}   &$$   \\
 \hline
 \multicolumn{2}{c|}{Channels}                &{Exp.}  &{This work}   &{Exp.}  &{This work}    &$R_{\perp}$   \\
 \hline
 1 &$B^- \to D^{*0} D^{*-}$                 &$$        &$0.2_{-0.1-0.1}^{+0.0+0.0}$       &$$              &$...$   &$0.17$\\
 2 &$B^- \to D^{*0} D_{s}^{*-}$             &$$        &$\sim -10^{-3}$    &$$              &$...$   &$0.16$\\
 3 &$\bar B^0 \to D^{*+} D^{*-}$            &$2\pm10$  &$\sim -10^{-2}$     &$-0.67\pm0.18$  &$-0.76_{-0.01-0.03-0.02}^{+0.00+0.03+0.02}$   &$0.16$\\
 4 &$\bar B^0 \to D^{*+} D_s^{*-}$          &$$        &$...$       &$$             &$...$   &$$\\
 5 &$\bar B_s^0 \to D_s^{*+} D^{*-}$        &$$        &$0.1_{-0.0-0.0}^{+0.1+0.1}$       &$$              &$...$   &$0.14$\\
 6 &$\bar B_s^0 \to D_s^{*+} D_s^{*-}$      &$$        &$...$       &$$               &$...$   &$0.17$\\
 \hline
 7 &$\bar B^0 \to D^{*0} \bar D^{*0}$        &$$  &$-3.4_{-0.5-1.8}^{+0.4+0.4}$ &$$               &$-0.73_{-0.05-0.14-0.01}^{+0.03+0.23+0.01}$   &$0.24$\\
 8 &$\bar B^0 \to D_s^{*+} D_s^{*-}$        &$$   &$-0.4_{-0.1-0.2}^{+0.1+0.3}$ &$$              &$-0.68_{-0.06-0.17-0.01}^{+0.03+0.27+0.01}$   &$0.32$\\
 9 &$\bar B_s^0 \to D^{*0}\bar D^{*0}$       &$$  &$0.2_{-0.1-0.0}^{+0.0+0.0}$  &$$              &$\sim 10^{-3}$   &$0.25$\\
 10 &$\bar B_s^0 \to D^{*+} D^{*-}$           &$$   &$...$&$$               &$...$   &$0.29$\\
 \hline\hline
 \end{tabular}
 \end{center}
 \end{table}


The first 6 channels in each of  Tables \ref{tab:BRforBtoPP},
\ref{tab:BRforBtoPV}, \ref{tab:BRforBtoVP}, \ref{tab:BRforBtoVV} and
\ref{tab:CPforBtoVV}, receive contributions from both emission
diagrams and annihilation diagrams; while the last 4 channels in
each table are pure annihilation processes. In order to make our
discussions easier, we give a number to each channel in the
beginning of each line in the tables.

 Compared with the tree operators, the penguin operators give very
small contributions because of the severe  suppression of the Wilson
coefficients. By calculating the ratio of the branching fraction
with only penguin contributions and that with all contributions in
the same channel, we estimate how much the penguin operators
contribute. Our results show that the penguin operators contribute
$0.1\%-0.2\%$ in those channels with a W emission contribution, and
contribute $0.3\%-0.7\%$ in those pure annihilation processes. Thus
it is enough to pay our attention only to the tree operators in the
following for the investigation of the branching ratios. Different
from the counting rules of the PQCD calculation of $B$ to two light
mesons decays, the nonfactorizable emission diagrams may give large
contributions because the asymmetry of the two quarks in charmed
mesons can not make the two diagrams nearly cancel each other.
However, from Eq.(\ref{eq:ampf})-Eq.(\ref{eq:ampl}), one can find
that the contributions of the nonfactorizable emission diagrams are
suppressed by the small Wilson coefficient $C_1$. Since the charm
quark is heavier than the u, d, s quark, the gluons in Fig.
\ref{fig:annihilationc} are softer than those in Fig.
\ref{fig:annihilation}. This indicates that the diagrams in Fig.
\ref{fig:annihilationc} will give larger contributions than those in
Fig. \ref{fig:annihilation}. It is confirmed by our numerical
results. However, these contributions are still much smaller than
those from the factorizable emission diagrams. Thus, the branching
ratios of the first 6 channels in Tables \ref{tab:BRforBtoPP},
\ref{tab:BRforBtoPV}, \ref{tab:BRforBtoVP}, and \ref{tab:BRforBtoVV}
are dominated by the factorizable emission diagrams.

Because the factorizable emission diagrams are dominate, the
amplitude of the first 6 channels in each table should be nearly
proportional to the product of a decay constant and a $B\to D$
transition form factor. Based on this physical picture, we can have
the following simple conclusions:

\begin{enumerate}
\item In each table the channels 1 and 3 should have similar
branching ratios because they have the same CKM matrix elements for
the factorizable emission diagrams and similar transition form
factors for isospin symmetry. For the same reason, channels 2 and 4
should also have similar branching ratios.

\item The branching ratios of channels 4 and 6 indicate that the $\bar
B_s\to D_s^{(*)}$ transition has a little smaller form factor than
$B\to D^{(*)}$ transition. The reason is that the antistrange quark
in the $\bar B_s$ meson has a little larger momentum fraction than
the d quark in the $\bar B^0$ meson due to the SU(3) breaking effect
\cite{bs}.  In \cite{Li:BtoDs0}, the $\bar B_s\to D_s$ transition is
investigated with the light-cone sum rules, and a similar branching
ratio for $\bar B_s\to D_s^+ D_s^-$ is obtained under the
factorization assumption. This means the PQCD and the light-cone sum
rules have the similar $\bar B_s \to D_s$ transition form factors.

\item The first 6 $B\to PP$ decays in Table \ref{tab:BRforBtoPP} and
the corresponding 6 $B \to PV$ decays in Table \ref{tab:BRforBtoPV}
have the same transition form factors, respectively, as well as the
similar decay constants between $D$ meson and $D^*$ meson. Thus
their branching ratios should also be similar. However, such
phenomena are not expected in $B\to VP$ decays and $B\to VV$ decays,
because in addition to the longitudinal polarization's
contributions, the $B \to VV$ decays also receive large
contributions from transverse polarizations.
\end{enumerate}

 From Tables \ref{tab:BRforBtoPP},  \ref{tab:BRforBtoPV},
\ref{tab:BRforBtoVP}, and \ref{tab:BRforBtoVV}, one can find that
most experimental branching ratios agree with our conclusions in the
above paragraphs very well within the errors.  The authors in
Refs.~\cite{TestFactorization1,TestFactorization2} also investigate
the decays of $B$ to double charmed mesons under factorization
assumption, but with different models. Their results also indicate
that the factorization works well.

Since the direct CP asymmetry is proportional to the interference
between the  tree and penguin contributions \cite{direct}, it should
be small because of the small penguin contributions as we mentioned
above. Our numerical results indeed indicate that the direct CP
violations are very small. A relatively large direct CP violation
appears in the pure annihilation decay $\bar B^0\to D^0 \bar D^0$
and its corresponding $B\to PV$, $VP$, and $VV$ decays. However it
is still only several percent. Although the experiments give somehow
large direct CP asymmetry in some channels, the uncertainty is still
large.  Any large direct CP violation observed in experiments would
be treated as a signal of new physics at first.

The mixing induced CP asymmetry in B decays is almost proportional
to the $\sin 2\beta$ from Eq.(\ref{2beta}), if we neglect the small
contribution from penguin contributions. It should be mentioned
that, experimentally the P wave component in the amplitudes of $B\to
VV$ mode will bring systematic errors in the results of mixing
induced CP asymmetry because they will bring a minus sign relative
to the S and D wave component. Our results for $A^{mix}_{CP}$ in
Table \ref{tab:CPforBtoVV} only include the S wave and D wave
contributions, and in this table we also give the values of
$R_{\perp}$, which is defined by the ratio of branching fractions
with only P wave component and that with all the contributions.
Because the P wave contributions are very small in the color allowed
tree dominated processes, the experimental measurements are still in
agreement with our calculations. For the pure annihilation
processes, the P-wave contributions are relatively large and
therefore these channels may not be good choices for the observation
of mixing induced CP asymmetry.

 In Table \ref{tab:BRforBtoVV}, we give the ratios of
transverse polarizations' contributions in branching ratios. One
can find that both in the processes with an external W emission
and in the pure annihilation decays, the transverse polarizations
take about $40\%-50\%$ of the contributions, which agree with the
present experimental data amazingly well. We should point out that
these ratios are very sensitive to the terms with power $r^2$
($r=m_D/m_B$), although these corrections change the other
observables only a little. With the $r^2$ corrections absent, the
ratios of transverse polarizations for the channels with an
external W emission are about $20\%$, and those for the pure
annihilation channels are $0$, because the transverse
polarization's contributions for these channels are at the power
of $r^2$. For the sensitivity of these ratios to the power
correction terms, we vary the variable $r$ by $20\%$ for an error
estimation in Table \ref{tab:BRforBtoVV}. In \cite{Li:BtoVV}, the
authors obtain the values for the ratios $\sim 50\%$ for the
external W emission processes simply by means of kinematics under
the naive factorization, which agree with our results. For the
pure annihilation decays, the transverse polarizations are
suppressed by $r^2$, which is the reason why the authors in
\cite{Li:BtoVV} think the ratios of the transverse polarizations
are very small. However, our calculation show that, with the $r^2$
terms included, these ratios increase to about $50\%$. This means
that the polarization fractions are quite sensitive to the power
corrections although they are not sensitive to the higher order
QCD corrections etc. The future experiments will tell us more
about the polarizations in the pure annihilation processes.


\section{Summary}
\label{Sec:summary}

Although the D meson mass is not very small compared with the B
meson mass, factorization can still work in the leading order of
$m_D/m_B$ and $\Lambda_{\rm{QCD}}/m_D$ expansion. Since the PQCD
approach can eliminate the end-point singularity in the perturbative
calculation, we investigate the decays of $B$ to double charmed
mesons systematically. Both pseudoscalar and vector charmed mesons
are included in the final states. We find that the factorizable
emission diagrams are dominant in the branching ratios.  Most of our
branching ratios agree with the experimental data, which means the
factorization assumption works well. However, experimental data show
that there are still some discrepancies, which means more work is
needed both at the theoretical side and the experimental side.

Our results indicate that the direct CP asymmetries in these
channels are very small. Thus, it will be a signal of new physics if
a large direct CP asymmetry appears. In the decays of $B$ to double
vector charmed mesons, the transverse polarizations contribute
$40\%-50\%$ both in the external W emission processes and in the
pure annihilation decays, which agree with the present experimental
data. We should mention that the correction terms at the power of
$r^2$ play an important role in transverse polarizations, without
which the ratios for the external W emission processes decrease to
about $20\%$ and for the pure annihilation decays the ratios are $0$
because of the $r^2$ suppression.

\section{acknowledgement}
This work is partly supported by National Natural Science Foundation
of China under Grants No. 10735080, No. 10625525, and No. 10525523.
We would like to thank W. Wang, Y.M. Wang, H. Zou, K. Ukai and
A.Satpath for valuable discussions.

\begin{appendix}

\section{scales and functions for the hard kernel}
\label{appendix:scale&h}

The variables that are used to determine the scales and the
expressions of the hard kernels are defined by
\begin{eqnarray}
 P_{en}&=&m_B^2 x_1 x_2 (1-r_3^2),\nonumber\\
 P_{en}^{(1)}&=&m_B^2 x_2  (x_1(1-r_3^2)-x_3(1-r_2^2-r_3^2)),\nonumber\\
 P_{en}^{(2)}&=&-m_B^2[x_2(x_3-1)r_2^2 +r_3^2(x_1(x_2-1)+x_2(x_3-1)-x_3)-x_2(x_1+x_3-1)],\nonumber\\
 P_{an}&=&m_B^2(1-(1-r_2^2)x_3-x_2(1-x_3(1-r_2^2)-(1-x_3)r_3^2)),\nonumber\\
 P_{an}^{(1)}&=&m_B^2(1+x_1x_2(1-r_3^2)-(1-r_2^2-r_3^2)x_2x_3),\nonumber\\
 P_{an}^{(2)}&=&m_B^2(-x_3r_2^2+x_1((r_3^2-1)x_2+1)+x_3+x_2((x_3-1)r_3^2+(r_2^2-1)x_3+1)-1),\nonumber\\
 P_{an}^{(c)}&=&m_B^2(1-r_2^2-r_3^2)x_2 x_3,\nonumber\\
 P_{an}^{(1c)}&=&m_B^2[x_1((r_3^2-1)x_2+1)-(r_2^2-1)x_3 +x_2((x_3-1)r_3^2+(r_2^2-1)x_3+1)],\nonumber\\
 P_{an}^{(2c)}&=&m_B^2x_2((r_2^2+r_3^2-1)x_3-(r_3^2-1)x_1).\label{eq:scaleV}
\end{eqnarray}
The scales are determined as
\begin{eqnarray}
 t_e^{(1)}&=&\mbox{max}\{\sqrt{x_2(1-r_3^2)}m_B f_{\rm{err}},1/b_1,1/b_2\},\nonumber\\
 t_e^{(2)}&=&\mbox{max}\{\sqrt{x_1(1-r_3^2)}m_B f_{\rm{err}},1/b_1,1/b_2\},\nonumber\\
 t_{en}^{(1)}&=&\mbox{max}\{\sqrt{|P_{en}|} f_{\rm{err}},\sqrt{|P_{en}^{(1)}|} f_{\rm{err}},1/b_1,1/b_3\},\nonumber\\
 t_{en}^{(2)}&=&\mbox{max}\{\sqrt{|P_{en}|} f_{\rm{err}},\sqrt{|P_{en}^{(2)}|} f_{\rm{err}},1/b_1,1/b_3\},\nonumber\\
 t_a^{(1)}&=&\mbox{max}\{\sqrt{1-(1-r_3^2)x_2}m_B f_{\rm{err}},1/b_2,1/b_3\},\nonumber\\
 t_a^{(2)}&=&\mbox{max}\{\sqrt{1-(1-r_2^2)x_3}m_B f_{\rm{err}},1/b_2,1/b_3\},\nonumber\\
 t_{an}^{(1)}&=&\mbox{max}\{\sqrt{|P_{an}|} f_{\rm{err}},\sqrt{|P_{an}^{(1)}|} f_{\rm{err}},1/b_1,1/b_2\},\nonumber\\
 t_{an}^{(2)}&=&\mbox{max}\{\sqrt{|P_{an}|} f_{\rm{err}},\sqrt{|P_{an}^{(2)}|} f_{\rm{err}},1/b_1,1/b_2\},\nonumber\\
 t_a^{(1c)}&=&\mbox{max}\{\sqrt{(1-r_2^2-r_3^2)x_3}m_B f_{\rm{err}},1/b_2,1/b_3\},\nonumber\\
 t_a^{(2c)}&=&\mbox{max}\{\sqrt{(1-r_2^2-r_3^2)x_2}m_B f_{\rm{err}},1/b_2,1/b_3\},\nonumber\\
 t_{an}^{(1c)}&=&\mbox{max}\{\sqrt{|P_{an}^{(c)}|} f_{\rm{err}},\sqrt{|P_{an}^{(1c)}|} f_{\rm{err}},1/b_1,1/b_2\},\nonumber\\
 t_{an}^{(2c)}&=&\mbox{max}\{\sqrt{|P_{an}^{(c)}|} f_{\rm{err}},\sqrt{|P_{an}^{(2c)}|} f_{\rm{err}},1/b_1,1/b_2\},
\end{eqnarray}
with $f_{\rm{err}}$ varies from $0.75$ to $1.25$ for an error
estimation.

The functions of the hard kernels that appear in the factorization
formulae are given by
\begin{eqnarray}
 h_e(x_1,x_2,b_1,b_2)&=&K_{0}\left(\sqrt{x_1x_2}m_Bb_1\right)
 \nonumber \\
 & &\times \left[\theta(b_1-b_2)K_0\left(\sqrt{x_2}m_B
 b_1\right)I_0\left(\sqrt{x_2}m_Bb_2\right)\right.
 \nonumber \\
 & &\left.+\theta(b_2-b_1)K_0\left(\sqrt{x_2}m_Bb_2\right)
 I_0\left(\sqrt{x_2}m_Bb_1\right)\right]\;,\nonumber\\
 h_a(x_2,x_3,b_2,b_3)&=&\left(i\frac{\pi}{2}\right)^2
 H_0^{(1)}\left(\sqrt{x_2x_3}m_Bb_2\right)
 \nonumber \\
 & &\times\left[\theta(b_2-b_3)
 H_0^{(1)}\left(\sqrt{x_3}m_Bb_2\right)
 J_0\left(\sqrt{x_3}m_Bb_3\right)\right.
 \nonumber \\
 & &\left.+\theta(b_3-b_2)H_0^{(1)}\left(\sqrt{x_3}m_Bb_3\right)
 J_0\left(\sqrt{x_3}m_Bb_2\right)\right]\;,\nonumber\\
 h^{(j)}_{en}&=&\left[\theta(b_1-b_3)K_0\left(\sqrt{P_{en}}
 b_1\right)I_0\left(\sqrt{P_{en}}b_3\right) +\theta(b_3-b_1)K_0\left(\sqrt{P_{en}} b_3\right)
 I_0\left(\sqrt{P_{en}} b_1\right)\right]
 \nonumber \\
 &  & \times \left\{ \begin{array}{cc}
 K_{0}(\sqrt{|P_{en}^{(j)}|}b_{3}) &  \mbox{for $P_{en}^{(j)} \geq 0$}  \\
 \frac{i\pi}{2} H_{0}^{(1)}(\sqrt{|P_{en}^{(j)}|}b_{3})  &
 \mbox{for $P_{en}^{(j)} \leq 0$}
  \end{array} \right\}\;,\nonumber\\
 h^{(j)}_{an}&=& i\frac{\pi}{2}
 \left[\theta(b_1-b_2)H_0^{(1)}\left(\sqrt{P_{an}}
 b_1\right)J_0\left(\sqrt{P_{an}}b_2\right) +\theta(b_2-b_1)H_0^{(1)}\left(\sqrt{P_{an}} b_2\right)
 J_0\left(\sqrt{P_{an}} b_1\right)\right]\;  \nonumber \\
 &  & \times \left\{ \begin{array}{cc}
 K_{0}(\sqrt{|P_{an}^{(j)}|}b_{1}) &  \mbox{for $P_{an}^{(j)} \geq 0$}  \\
 \frac{i\pi}{2} H_{0}^{(1)}(\sqrt{|P_{an}^{(j)}|}b_{1})  &
 \mbox{for $P_{an}^{(j)} \leq 0$}
  \end{array} \right\}\;,
\end{eqnarray}
and the functions that consist of coupling constant and Sudakov
factors are given by
 \begin{eqnarray}
 E_e(t)&=&\alpha_s(t)\exp[-S_B(t)-S_{M_2}(t)]\;,\nonumber \\
 E_a(t)&=&\alpha_s(t)\exp[-S_{M_2}(t)-S_{M_3}(t)]\;,\nonumber\\
 E_{en}(t)&=&\alpha_s(t)\exp[-S_B(t)-S_{M_2}-S_{M_3}|_{b_2=b_1}]\;,\nonumber \\
 E_{an}(t)&=&\alpha_s(t)\exp[-S_B(t)-S_{M_2}-S_{M_3}|_{b_3=b_2}]\;,
 \end{eqnarray}
where
 \begin{eqnarray}
 S_B(t)=S_{M_2}=S_{M_3}=s\left(x_1\frac{m_{B}}{\sqrt{2}},b_1\right)+\frac{5}{3}\int^t_{1/b_1}\frac{d\bar \mu}{\bar
 \mu}\gamma_q(\alpha_s(\bar \mu)),
 \end{eqnarray}
 with the quark anomalous dimension $\gamma_q=-\alpha_s/\pi$. The
explicit form for the  function $s(Q,b)$ is:
\begin{eqnarray}
s(Q,b)&=&~~\frac{A^{(1)}}{2\beta_{1}}\hat{q}\ln\left(\frac{\hat{q}}
{\hat{b}}\right)-
\frac{A^{(1)}}{2\beta_{1}}\left(\hat{q}-\hat{b}\right)+
\frac{A^{(2)}}{4\beta_{1}^{2}}\left(\frac{\hat{q}}{\hat{b}}-1\right)
-\left[\frac{A^{(2)}}{4\beta_{1}^{2}}-\frac{A^{(1)}}{4\beta_{1}}
\ln\left(\frac{e^{2\gamma_E-1}}{2}\right)\right]
\ln\left(\frac{\hat{q}}{\hat{b}}\right)
\nonumber \\
&&+\frac{A^{(1)}\beta_{2}}{4\beta_{1}^{3}}\hat{q}\left[
\frac{\ln(2\hat{q})+1}{\hat{q}}-\frac{\ln(2\hat{b})+1}{\hat{b}}\right]
+\frac{A^{(1)}\beta_{2}}{8\beta_{1}^{3}}\left[
\ln^{2}(2\hat{q})-\ln^{2}(2\hat{b})\right],
\end{eqnarray} where the variables are defined by
\begin{eqnarray}
\hat q\equiv \mbox{ln}[Q/(\sqrt 2\Lambda)],~~~ \hat b\equiv
\mbox{ln}[1/(b\Lambda)], \end{eqnarray} and the coefficients
$A^{(i)}$ and $\beta_i$ are \begin{eqnarray}
\beta_1=\frac{33-2n_f}{12},~~\beta_2=\frac{153-19n_f}{24},\nonumber\\
A^{(1)}=\frac{4}{3},~~A^{(2)}=\frac{67}{9}
-\frac{\pi^2}{3}-\frac{10}{27}n_f+\frac{8}{3}\beta_1\mbox{ln}(\frac{1}{2}e^{\gamma_E}),
\end{eqnarray}
$n_f$ is the number of the quark flavors and $\gamma_E$ is the Euler
constant. We will use the one-loop running coupling constant, i.e.
we pick up the four terms in the first line of the expression for
the function $s(Q,b)$.

\section{factorization formulae for $B \to PV$($M_2$ is a pseudoscalar meson and $M_3$ is a vector meson)}
\label{appendix:BtoPV}

\begin{eqnarray}
 F_e^{LL}(a_i(t))&=&8\pi C_Ff_{M_3}m_B^4 \int_{0}^{1}d x_{1}d
 x_{2}\int_{0}^{1/\Lambda} b_1d b_1 b_2d b_2
 \phi_B(x_1,b_1)\phi_{M_2}(x_2)\nonumber \\
 &&\times \left[-\big((2x_2-1)r_2-x_2-1\big)E_e(t_e^{(1)})a_i(t_e^{(1)})h_e(x_1,x_2(1-r_3^2),b_1,b_2)S_t(x_2)\right.\nonumber\\
 &&\left.+r_2(1+r_2)E_e(t_e^{(2)})a_i(t_e^{(2)})h_e(x_2,x_1(1-r_3^2),b_2,b_1)S_t(x_1)\right]\;,\\
 F_e^{SP}(a_i(t))&=&0\;,
 \end{eqnarray}
 \begin{eqnarray}
 F_{en}^{LL}(a_i(t))&=& 16\pi\sqrt{\frac{2}{3}} C_F m_B^4 \int_0^1
 [dx]\int_0^{1/\Lambda} b_1 db_1 b_3 db_3
 \phi_B(x_1,b_1)\phi_{M_2}(x_2)\phi_{M_3}(x_3) \nonumber \\
 & &\times
 \left[\big(-x_2r_2+x_3\big)E_b(t_{en}^{(1)})a_i(t_{en}^{(1)})h^{(1)}_{en}(x_i,b_i)\right.\nonumber\\
 &&\left.+\big(x_2r_2-x_2+x_3-1\big)E_{en}(t_{en}^{(2)})a_i(t_{en}^{(2)})h^{(2)}_{en}(x_i,b_i) \right]\;,
 \end{eqnarray}
 \begin{eqnarray}
 F_{en}^{LR}(a_i(t))&=& 16\pi\sqrt{\frac{2}{3}} C_F m_B^4 \int_0^1
 [dx]\int_0^{1/\Lambda} b_1 db_1 b_3 db_3
 \phi_B(x_1,b_1)\phi_{M_2}(x_2)\phi_{M_3}(x_3)
 \nonumber \\
 & &\times r_3 \left[\big(x_3-r_2(x_2-x_3)\big)E_{en}(t_{en}^{(1)})a_i(t_{en}^{(1)})h^{(1)}_{en}(x_i,b_i)\right.\nonumber\\
 &&\left.-\big(x_3+r_2(x_2+x_3)\big)E_{en}(t_{en}^{(2)})a_i(t_{en}^{(2)})h^{(2)}_{en}(x_i,b_i) \right]\;,
 \end{eqnarray}
 \begin{eqnarray}
 F_a^{LL}(a_i(t))&=&8\pi C_Ff_B m_B^4 \int_0^1
 dx_2dx_3\int_0^{1/\Lambda}b_2db_2b_3db_3 \phi_{M_2}(x_2) \phi_{M_3}(x_3)\nonumber \\
 & &\times \left[\big(x_2-1\big)E_a(t_a^{(1)})a_i(t_a^{(1)}) h_a(1-(1-r_2^2)x_3,1-(1-r_3^2)x_2,b_3,b_2)S_t(x_2)\right.\nonumber \\
 & &\left.+\big(-2r_2r_3x_3-(x_3-1)\big) E_a(t_a^{(2)})a_i(t_a^{(2)}) h_a(1-(1-r_3^2)x_2,1-(1-r_2^2)x_3,b_2,b_3)S_t(x_3)\right]\;,\\
 F_a^{LR}(a_i(t))&=&-F_a^{LL}(a_i(t)),
 \end{eqnarray}
 \begin{eqnarray}
 F_a^{SP}(a_i(t))&=&16\pi C_Ff_B m_B^4 \int_0^1
 dx_2dx_3\int_0^{1/\Lambda}b_2db_2b_3db_3 \phi_{M_2}(x_2) \phi_{M_3}(x_3)\nonumber \\
 & &\times \left[r_2(1-x_2)E_a(t_a^{(1)})a_i(t_a^{(1)}) h_a(1-(1-r_2^2)x_3,1-(1-r_3^2)x_2,b_3,b_2)S_t(x_2)\right.\nonumber \\
 & &\left.+(2r_2+r_3(x_3-1))E_a(t_a^{(2)})a_i(t_a^{(2)}) h_a(1-(1-r_3^2)x_2,1-(1-r_2^2)x_3,b_2,b_3)S_t(x_3)\right]\;,
 \end{eqnarray}
 \begin{eqnarray}
 F_{ac}^{LL}(a_i(t))&=&8\pi C_Ff_B m_B^4 \int_0^1
 dx_2dx_3\int_0^{1/\Lambda}b_2db_2b_3db_3 \phi_{M_2}(x_2) \phi_{M_3}(x_3)\nonumber \\
 & &\times \left[\big(r_2r_3(1-2x_3)-x_3\big)E_a(t_a^{(1c)})a_i(t_a^{(1c)}) h_a(x_2,x_3(1-r_2^2-r_3^2),b_2,b_3)S_t(x_3)\right.\nonumber \\
 &&\left.+\big(r_2r_3+x_2\big)E_a(t_a^{(2c)})a_i(t_a^{(2c)})h_a(x_3,x_2(1-r_2^2-r_3^2),b_3,b_2)S_t(x_2)\right]\;,\\
 F_{ac}^{LR}(a_i(t))&=&-F_{ac}^{LL}(a_i(t)),
 \end{eqnarray}
 \begin{eqnarray}
 F_{an}^{LL}(a_i(t))&=& 16\pi\sqrt{\frac{2}{3}} C_F m_B^4 \int_0^1
 [dx]\int_0^{1/\Lambda}b_1 db_1 b_2db_2\phi_B(x_1,b_1)\phi_{M_2}(x_2)\phi_{M_3}(x_3) \nonumber \\
 & &\times \left[\big(r_2r_3(x_3-x_2)+x_3-1\big)
 E_{an}(t_{an}^{(1)})a_i(t_{an}^{(1)})h^{(1)}_{an}(x_i,b_i)\right.\nonumber\\
 &&\left.+\big(r_2r_3(x_3-x_2)-x_2+1\big) E_{an}(t_{an}^{(2)})a_i(t_{an}^{(2)})h^{(2)}_{an}(x_i,b_i) \right]\;,
 \end{eqnarray}
 \begin{eqnarray}
 F_{an}^{LR}(a_i(t))&=& 16\pi\sqrt{\frac{2}{3}} C_F m_B^4 \int_0^1
 [dx]\int_0^{1/\Lambda}b_1 db_1 b_2db_2\phi_B(x_1,b_1)\phi_{M_2}(x_2)\phi_{M_3}(x_3) \nonumber \\
 & &\times \left[-\big(r_2(x_2+1)-r_3(x_3+1)\big) E_{an}(t_{an}^{(1)})a_i(t_{an}^{(1)})h^{(1)}_{an}(x_i,b_i)\right.\nonumber\\
 &&\left.+ (r_2(x_2-1)-r_3(x_3-1))E_{an}(t_{an}^{(2)})a_i(t_{an}^{(2)})h^{(2)}_{an}(x_i,b_i) \right]\;,
 \end{eqnarray}
 \begin{eqnarray}
 F_{an}^{SP}(a_i(t))&=& 16\pi\sqrt{\frac{2}{3}} C_F m_B^4 \int_0^1
 [dx]\int_0^{1/\Lambda}b_1 db_1 b_2db_2\phi_B(x_1,b_1)\phi_{M_2}(x_2)\phi_{M_3}(x_3) \nonumber \\
 & &\times \left[-\big(r_2r_3(x_2-x_3)+x_2-1\big) E_{an}(t_{an}^{(1)})a_i(t_{an}^{(1)})h^{(1)}_{an}(x_i,b_i)\right.\nonumber\\
 &&\left.-\big(r_2r_3(x_2-x_3)-x_3+1\big)E_{an}(t_{an}^{(2)})a_i(t_{an}^{(2)})h^{(2)}_{an}(x_i,b_i) \right]\;,
 \end{eqnarray}
 \begin{eqnarray}
 F_{anc}^{LL}(a_i(t))&=& 16\pi\sqrt{\frac{2}{3}} C_F m_B^4 \int_0^1
 [dx]\int_0^{1/\Lambda}b_1 db_1 b_2db_2\phi_B(x_1,b_1)\phi_{M_2}(x_2)\phi_{M_3}(x_3) \nonumber \\
 & &\times \left[\big(r_2r_3(x_3-x_2)-x_2\big) E_{an}(t_{an}^{(1c)})a_i(t_{an}^{(1c)})h^{(1c)}_{an}(x_i,b_i)\right.\nonumber\\
 &&\left.+\big(r_2r_3(x_3-x_2)+x_3\big) E_{an}(t_{an}^{(2c)})a_i(t_{an}^{(2c)})h^{(2c)}_{an}(x_i,b_i) \right]\;,
 \end{eqnarray}
 \begin{eqnarray}
 F_{anc}^{SP}(a_i(t))&=& 16\pi\sqrt{\frac{2}{3}} C_F m_B^4 \int_0^1
 [dx]\int_0^{1/\Lambda}b_1 db_1 b_2db_2\phi_B(x_1,b_1)\phi_{M_2}(x_2)\phi_{M_3}(x_3) \nonumber \\
 & &\times \left[\big(r_2r_3(x_3-x_2)+x_3\big) E_{an}(t_{an}^{(1c)})a_i(t_{an}^{(1c)})h^{(1c)}_{an}(x_i,b_i)\right.\nonumber\\
 &&\left.-\big((r_2r_3+1)x_2-r_3x_3(r_2+r_3)\big) E_{an}(t_{an}^{(2c)})a_i(t_{an}^{(2c)})h^{(2c)}_{an}(x_i,b_i) \right]\;,
\end{eqnarray}

\section{factorization formulae for $B \to VP$($M_2$ is a vector meson and $M_3$ is a pseudoscalar meson)}
\label{appendix:BtoVP}

\begin{eqnarray}
 F_e^{LL}(a_i(t))&=&8\pi C_Ff_{M_3}m_B^4 \int_{0}^{1}d x_{1}d
 x_{2}\int_{0}^{1/\Lambda} b_1d b_1 b_2d b_2
 \phi_B(x_1,b_1)\phi_{M_2}(x_2)
 \nonumber \\
 &&\times \left[-\big((2x_2-1)r_2-x_2-1\big)E_e(t_e^{(1)})a_i(t_e^{(1)})h_e(x_1,x_2(1-r_3^2),b_1,b_2)S_t(x_2)\right.\nonumber\\
 &&\left.+r_2(1+r_2)E_e(t_e^{(2)})a_i(t_e^{(2)})h_e(x_2,x_1(1-r_3^2),b_2,b_1)S_t(x_1)\right]\;,
 \end{eqnarray}
 \begin{eqnarray}
 F_e^{SP}(a_i(t))&=&16\pi C_Ff_{M_3}m_B^4 \int_{0}^{1}d x_{1}d
 x_{2}\int_{0}^{1/\Lambda} b_1d b_1 b_2d b_2
 \phi_B(x_1,b_1)\phi_{M_2}(x_2)\nonumber \\
 & &\times r_3 \left[(r_2x_2-1) E_e(t_e^{(1)})a_i(t_e^{(1)})h_e(x_1,x_2(1-r_3^2),b_1,b_2)S_t(x_2)\right.\nonumber\\
 &&\left.-r_2E_e(t_e^{(2)})a_i(t_e^{(2)})h_e(x_2,x_1(1-r_3^2),b_2,b_1)S_t(x_1)\right]\;,
 \end{eqnarray}
 \begin{eqnarray}
 F_{en}^{LL}(a_i(t))&=& 16\pi\sqrt{\frac{2}{3}} C_F m_B^4 \int_0^1
 [dx]\int_0^{1/\Lambda} b_1 db_1 b_3 db_3
 \phi_B(x_1,b_1)\phi_{M_2}(x_2)\phi_{M_3}(x_3)
 \nonumber \\
 & &\times \left[-\big(-x_2r_2-x_3\big)
 E_b(t_{en}^{(1)})a_i(t_{en}^{(1)})h^{(1)}_{en}(x_i,b_i)\right.\nonumber\\
 &&\left.+\big(x_2r_2-x_2+x_3-1\big)E_{en}(t_{en}^{(2)})a_i(t_{en}^{(2)})h^{(2)}_{en}(x_i,b_i) \right]\;,
 \end{eqnarray}
 \begin{eqnarray}
 F_{en}^{LR}(a_i(t))&=& 16\pi\sqrt{\frac{2}{3}} C_F m_B^4 \int_0^1
 [dx]\int_0^{1/\Lambda} b_1 db_1 b_3 db_3
 \phi_B(x_1,b_1)\phi_{M_2}(x_2)\phi_{M_3}(x_3)
 \nonumber \\
 & &\times r_3 \left[-\big(r_2(x_2-x_3)+x_3\big) E_{en}(t_{en}^{(1)})a_i(t_{en}^{(1)})h^{(1)}_{en}(x_i,b_i)\right.\nonumber\\
 &&\left.+\big(r_2(x_2+x_3-2)-x_3+2\big)E_{en}(t_{en}^{(2)})a_i(t_{en}^{(2)})h^{(2)}_{en}(x_i,b_i) \right]\;,
 \end{eqnarray}
 \begin{eqnarray}
 F_a^{LL}(a_i(t))&=&8\pi C_Ff_B m_B^4 \int_0^1
 dx_2dx_3\int_0^{1/\Lambda}b_2db_2b_3db_3 \phi_{M_2}(x_2) \phi_{M_3}(x_3)\nonumber \\
 & &\times \left[\big(2r_2r_3x_2+x_2-1\big)\right.\nonumber\\
 &&\left.\times E_a(t_a^{(1)})a_i(t_a^{(1)}) h_a(1-(1-r_2^2)x_3,1-(1-r_3^2)x_2,b_3,b_2)S_t(x_2)\right.\nonumber \\
 & &\left.+\big(1-x_3\big)E_a(t_a^{(2)})a_i(t_a^{(2)}) h_a(1-(1-r_3^2)x_2,1-(1-r_2^2)x_3,b_2,b_3)S_t(x_3)\right]\;,\\
 F_a^{LR}(a_i(t))&=&-F_a^{LL}(a_i(t)),
 \end{eqnarray}
 \begin{eqnarray}
 F_a^{SP}(a_i(t))&=&16\pi C_Ff_B m_B^4 \int_0^1
 dx_2dx_3\int_0^{1/\Lambda}b_2db_2b_3db_3 \phi_{M_2}(x_2) \phi_{M_3}(x_3)\nonumber \\
 & &\times \left[-\big(2r_3+r_2(x_2-1)\big)E_a(t_a^{(1)})a_i(t_a^{(1)}) h_a(1-(1-r_2^2)x_3,1-(1-r_3^2)x_2,b_3,b_2)S_t(x_2)\right.\nonumber \\
 & &\left.+r_3(x_3-1)E_a(t_a^{(2)})a_i(t_a^{(2)}) h_a(1-(1-r_3^2)x_2,1-(1-r_2^2)x_3,b_2,b_3)S_t(x_3)\right]\;,
\end{eqnarray}

\begin{eqnarray}
 F_{ac}^{LL}(a_i(t))&=&8\pi C_Ff_B m_B^4 \int_0^1
 dx_2dx_3\int_0^{1/\Lambda}b_2db_2b_3db_3 \phi_{M_2}(x_2) \phi_{M_3}(x_3)\nonumber \\
 & &\times \left[\big(-r_2r_3-x_3\big)E_a(t_a^{(1c)})a_i(t_a^{(1c)}) h_a(x_2,x_3(1-r_2^2-r_3^2),b_2,b_3)S_t(x_3)\right.\nonumber \\
 & &\left.-\big(r_2r_3(1-2x_2)-x_2\big) a_i(t_a^{(2c)}) h_a(x_3,x_2(1-r_2^2-r_3^2),b_3,b_2)S_t(x_2)\right]\;,\\
 F_{ac}^{LR}(a_i(t))&=&-F_{ac}^{LL}(a_i(t)),
 \end{eqnarray}
 \begin{eqnarray}
 F_{an}^{LL}(a_i(t))&=& 16\pi\sqrt{\frac{2}{3}} C_F m_B^4 \int_0^1
 [dx]\int_0^{1/\Lambda}b_1 db_1 b_2db_2\phi_B(x_1,b_1)\phi_{M_2}(x_2)\phi_{M_3}(x_3) \nonumber \\
 & &\times \left[-\big(r_2r_3(x_2-x_3)-x_3+1\big)
 E_{an}(t_{an}^{(1)})a_i(t_{an}^{(1)})h^{(1)}_{an}(x_i,b_i)\right.\nonumber\\
 &&\left.+\big(r_2r_3(x_3-x_2)-x_2+1\big) E_{an}(t_{an}^{(2)})a_i(t_{an}^{(2)})h^{(2)}_{an}(x_i,b_i) \right]\;,
 \end{eqnarray}
 \begin{eqnarray}
 F_{an}^{LR}(a_i(t))&=& 16\pi\sqrt{\frac{2}{3}} C_F m_B^4 \int_0^1
 [dx]\int_0^{1/\Lambda}b_1 db_1 b_2db_2\phi_B(x_1,b_1)\phi_{M_2}(x_2)\phi_{M_3}(x_3) \nonumber \\
 & &\times \left[\big(r_2(x_2+1)-r_3(x_3+1)\big) E_{an}(t_{an}^{(1)})a_i(t_{an}^{(1)})h^{(1)}_{an}(x_i,b_i)\right.\nonumber\\
 &&\left.-\big(r_2(x_2-1)-r_3(x_3-1)\big) E_{an}(t_{an}^{(2)})a_i(t_{an}^{(2)})h^{(2)}_{an}(x_i,b_i) \right]\;,
 \end{eqnarray}
 \begin{eqnarray}
 F_{an}^{SP}(a_i(t))&=& 16\pi\sqrt{\frac{2}{3}} C_F m_B^4 \int_0^1
 [dx]\int_0^{1/\Lambda}b_1 db_1 b_2db_2\phi_B(x_1,b_1)\phi_{M_2}(x_2)\phi_{M_3}(x_3) \nonumber \\
 & &\times \left[\big(r_2r_3(x_3-x_2)-x_2+1\big) E_{an}(t_{an}^{(1)})a_i(t_{an}^{(1)})h^{(1)}_{an}(x_i,b_i)\right.\nonumber\\
 &&\left.- \big(r_2r_3(x_2-x_3)-x_3+1\big) E_{an}(t_{an}^{(2)})a_i(t_{an}^{(2)})h^{(2)}_{an}(x_i,b_i) \right]\;,
 \end{eqnarray}
 \begin{eqnarray}
 F_{anc}^{LL}(a_i(t))&=& 16\pi\sqrt{\frac{2}{3}} C_F m_B^4 \int_0^1
 [dx]\int_0^{1/\Lambda}b_1 db_1 b_2db_2\phi_B(x_1,b_1)\phi_{M_2}(x_2)\phi_{M_3}(x_3) \nonumber \\
 & &\times \left[\big(r_2r_3(x_3-x_2)-x_2\big) E_{an}(t_{an}^{(1c)})a_i(t_{an}^{(1c)})h^{(1c)}_{an}(x_i,b_i)\right.\nonumber\\
 &&\left.-\big(r_2r_3(x_2-x_3)-x_3\big) E_{an}(t_{an}^{(2c)})a_i(t_{an}^{(2c)})h^{(2c)}_{an}(x_i,b_i) \right]\;,
 \end{eqnarray}
 \begin{eqnarray}
 F_{anc}^{SP}(a_i(t))&=& 16\pi\sqrt{\frac{2}{3}} C_F m_B^4 \int_0^1
 [dx]\int_0^{1/\Lambda}b_1 db_1 b_2db_2\phi_B(x_1,b_1)\phi_{M_2}(x_2)\phi_{M_3}(x_3) \nonumber \\
 & &\times \left[-\big(r_2r_3(x_2-x_3)-x_3\big) E_{an}(t_{an}^{(1c)})a_i(t_{an}^{(1c)})h^{(1c)}_{an}(x_i,b_i)\right.\nonumber\\
 &&\left.\big((-r_2r_3-1)x_2+r_2r_3x_3\big) E_{an}(t_{an}^{(2c)})a_i(t_{an}^{(2c)})h^{(2c)}_{an}(x_i,b_i) \right]\;,
\end{eqnarray}

\section{factorization formulae for $B \to VV$}
\label{appendix:BtoVV}

\subsection{Longitudinal polarization}
\begin{eqnarray}
 F_e^{LL}(a_i(t))&=&8\pi C_Ff_{M_3}m_B^4 \int_{0}^{1}d x_{1}d
 x_{2}\int_{0}^{1/\Lambda} b_1d b_1 b_2d b_2
 \phi_B(x_1,b_1)\phi_{M_2}(x_2)\nonumber \\
 &&\times \left[(-1-x_2+r_2(2x_2-1))E_e(t_e^{(1)})a_i(t_e^{(1)})h_e(x_1,x_2(1-r_3^2),b_1,b_2)S_t(x_2)\right.\nonumber\\
 &&\left.-r_2(1+r_2)E_e(t_e^{(2)})a_i(t_e^{(2)})h_e(x_2,x_1(1-r_3^2),b_2,b_1)S_t(x_1)\right]\;,\\
 F_e^{SP}(a_i(t))&=&0\;,
 \end{eqnarray}
 \begin{eqnarray}
 F_{en}^{LL}(a_i(t))&=& 16\pi\sqrt{\frac{2}{3}} C_F m_B^4 \int_0^1
 [dx]\int_0^{1/\Lambda} b_1 db_1 b_3 db_3
 \phi_B(x_1,b_1)\phi_{M_2}(x_2)\phi_{M_3}(x_3)\nonumber \\
 & &\times
 \left[(-x_2r_2-x_3)E_b(t_{en}^{(1)})a_i(t_{en}^{(1)})h^{(1)}_{en}(x_i,b_i)\right.\nonumber\\
 &&\left.-\big(x_2r_2-x_2+x_3-1\big)E_{en}(t_{en}^{(2)})a_i(t_{en}^{(2)})h^{(2)}_{en}(x_i,b_i) \right]\;,
 \end{eqnarray}
 \begin{eqnarray}
 F_{en}^{LR}(a_i(t))&=& 16\pi\sqrt{\frac{2}{3}} C_F m_B^4 \int_0^1
 [dx]\int_0^{1/\Lambda} b_1 db_1 b_3 db_3
 \phi_B(x_1,b_1)\phi_{M_2}(x_2)\phi_{M_3}(x_3)\nonumber \\
 & &\times r_3 \left[-(r_2(x_3+x_2)-x_3)E_{en}(t_{en}^{(1)})a_i(t_{en}^{(1)})h^{(1)}_{en}(x_i,b_i)\right.\nonumber\\
 &&\left.-\big(x_3+r_2(x_2-x_3)\big)E_{en}(t_{en}^{(2)})a_i(t_{en}^{(2)})h^{(2)}_{en}(x_i,b_i) \right]\;,
 \end{eqnarray}
 \begin{eqnarray}
 F_a^{LL}(a_i(t))&=&8\pi C_Ff_B m_B^4 \int_0^1
 dx_2dx_3\int_0^{1/\Lambda}b_2db_2b_3db_3 \phi_{M_2}(x_2) \phi_{M_3}(x_3)\nonumber \\
 & &\times \left[(-x_2+1)E_a(t_a^{(1)})a_i(t_a^{(1)}) h_a(1-(1-r_2^2)x_3,1-(1-r_3^2)x_2,b_3,b_2)S_t(x_2)\right.\nonumber \\
 & &\left.-(1-x_3)E_a(t_a^{(2)})a_i(t_a^{(2)}) h_a(1-(1-r_3^2)x_2,1-(1-r_2^2)x_3,b_2,b_3)S_t(x_3)\right]\;,\\
 F_a^{LR}(a_i(t))&=&F_a^{LL}(a_i(t)),
 \end{eqnarray}
 \begin{eqnarray}
 F_a^{SP}(a_i(t))&=&16\pi C_Ff_B m_B^4 \int_0^1
 dx_2dx_3\int_0^{1/\Lambda}b_2db_2b_3db_3 \phi_{M_2}(x_2) \phi_{M_3}(x_3)\nonumber \\
 & &\times \left[r_2(x_2-1)E_a(t_a^{(1)})a_i(t_a^{(1)}) h_a(1-(1-r_2^2)x_3,1-(1-r_3^2)x_2,b_3,b_2)S_t(x_2)\right.\nonumber \\
 & &\left.+r_3(x_3-1)E_a(t_a^{(2)})a_i(t_a^{(2)}) h_a(1-(1-r_3^2)x_2,1-(1-r_2^2)x_3,b_2,b_3)S_t(x_3)\right]\;,
\end{eqnarray}
\begin{eqnarray}
 F_{ac}^{LL}(a_i(t))&=&0\;,\\
 F_{ac}^{LR}(a_i(t))&=&F_{ac}^{LL}(a_i(t))=0,\\
 F_{an}^{LL}(a_i(t))&=& 16\pi\sqrt{\frac{2}{3}} C_F m_B^4 \int_0^1
 [dx]\int_0^{1/\Lambda}b_1 db_1 b_2db_2\phi_B(x_1,b_1)\phi_{M_2}(x_2)\phi_{M_3}(x_3) \nonumber \\
 & &\times \left[\big(-r_3r_2(x_2+x_3)-x_3+1\big)
 E_{an}(t_{an}^{(1)})a_i(t_{an}^{(1)})h^{(1)}_{an}(x_i,b_i)\right.\nonumber\\
 &&\left.-\big(-r_2r_3(x_2+x_3-2)-x_2+1\big)
 E_{an}(t_{an}^{(2)})a_i(t_{an}^{(2)})h^{(2)}_{an}(x_i,b_i) \right]\;,
 \end{eqnarray}
 \begin{eqnarray}
 F_{an}^{LR}(a_i(t))&=& 16\pi\sqrt{\frac{2}{3}} C_F m_B^4 \int_0^1
 [dx]\int_0^{1/\Lambda}b_1 db_1 b_2db_2\phi_B(x_1,b_1)\phi_{M_2}(x_2)\phi_{M_3}(x_3) \nonumber \\
 & &\times \left[-(-r_3(1+x_3)+r_2(1+x_2)) E_{an}(t_{an}^{(1)})a_i(t_{an}^{(1)})h^{(1)}_{an}(x_i,b_i)\right.\nonumber\\
 &&\left.+ (r_2(x_2-1)-r_3(x_3-1))E_{an}(t_{an}^{(2)})a_i(t_{an}^{(2)})h^{(2)}_{an}(x_i,b_i) \right]\;,
 \end{eqnarray}
 \begin{eqnarray}
 F_{an}^{SP}(a_i(t))&=& 16\pi\sqrt{\frac{2}{3}} C_F m_B^4 \int_0^1
 [dx]\int_0^{1/\Lambda}b_1 db_1 b_2db_2\phi_B(x_1,b_1)\phi_{M_2}(x_2)\phi_{M_3}(x_3) \nonumber \\
 & &\times \left[\big(-r_2r_3(x_2+x_3)-x_2+1\big) E_{an}(t_{an}^{(1)})a_i(t_{an}^{(1)})h^{(1)}_{an}(x_i,b_i)\right.\nonumber\\
 &&\left.-\big(-r_2r_3(x_2+x_3-2)-(x_3-1)\big)E_{an}(t_{an}^{(2)})a_i(t_{an}^{(2)})h^{(2)}_{an}(x_i,b_i) \right]\;,
 \end{eqnarray}
 \begin{eqnarray}
 F_{anc}^{LL}(a_i(t))&=& 16\pi\sqrt{\frac{2}{3}} C_F m_B^4 \int_0^1
 [dx]\int_0^{1/\Lambda}b_1 db_1 b_2db_2\phi_B(x_1,b_1)\phi_{M_2}(x_2)\phi_{M_3}(x_3) \nonumber \\
 & &\times \left[-\big(-r_2r_3(x_2+x_3-2)-x_2\big) E_{an}(t_{an}^{(1c)})a_i(t_{an}^{(1c)})h^{(1c)}_{an}(x_i,b_i)\right.\nonumber\\
 &&\left.+\big(-r_2r_3(x_2+x_3)-x_3\big) E_{an}(t_{an}^{(2c)})a_i(t_{an}^{(2c)})h^{(2c)}_{an}(x_i,b_i) \right]\;,
 \end{eqnarray}
 \begin{eqnarray}
 F_{anc}^{SP}(a_i(t))&=& 16\pi\sqrt{\frac{2}{3}} C_F m_B^4 \int_0^1
 [dx]\int_0^{1/\Lambda}b_1 db_1 b_2db_2\phi_B(x_1,b_1)\phi_{M_2}(x_2)\phi_{M_3}(x_3) \nonumber \\
 & &\times \left[-\big(-r_2r_3(x_2+x_3-2)-x_3\big) E_{an}(t_{an}^{(1c)})a_i(t_{an}^{(1c)})h^{(1c)}_{an}(x_i,b_i)\right.\nonumber\\
 &&\left.+\big((-r_2r_3-1)x_2-r_3r_2x_3\big) E_{an}(t_{an}^{(2c)})a_i(t_{an}^{(2c)})h^{(2c)}_{an}(x_i,b_i) \right]\;,
\end{eqnarray}

\subsection{Transverse Polarization}

\begin{eqnarray}
 F_e^{LL,s}(a_i(t))&=&8\pi C_Ff_{M_3}m_B^4 \int_{0}^{1}d x_{1}d
 x_{2}\int_{0}^{1/\Lambda} b_1d b_1 b_2d b_2
 \phi_B(x_1,b_1)\phi_{M_2}(x_2)\nonumber \\
 &&\times r_3 \left[-\big(r_2(x_2+2)+1\big)E_e(t_e^{(1)})a_i(t_e^{(1)})h_e(x_1,x_2(1-r_3^2),b_1,b_2)S_t(x_2)\right.\nonumber\\
 &&\left.-r_2E_e(t_e^{(2)})a_i(t_e^{(2)})h_e(x_2,x_1(1-r_3^2),b_2,b_1)S_t(x_1)\right]\;,
 \end{eqnarray}
 \begin{eqnarray}
 F_e^{LL,p}(a_i(t))&=&8\pi C_Ff_{M_3}m_B^4 \int_{0}^{1}d x_{1}d
 x_{2}\int_{0}^{1/\Lambda} b_1d b_1 b_2d b_2
 \phi_B(x_1,b_1)\phi_{M_2}(x_2)\nonumber \\
 &&\times r_3 \left[(r_2x_2-1)E_e(t_e^{(1)})a_i(t_e^{(1)})h_e(x_1,x_2(1-r_3^2),b_1,b_2)S_t(x_2)\right.\nonumber\\
 &&\left.-r_2E_e(t_e^{(2)})a_i(t_e^{(2)})h_e(x_2,x_1(1-r_3^2),b_2,b_1)S_t(x_1)\right]\;,\\
 F_e^{SP,s}(a_i(t))&=&F_e^{SP,p}(a_i(t))=0\;,
 \end{eqnarray}
 \begin{eqnarray}
 F_{en}^{LL,s}(a_i(t))&=& 16\pi\sqrt{\frac{2}{3}} C_F m_B^4 \int_0^1
 [dx]\int_0^{1/\Lambda} b_1 db_1 b_3 db_3
 \phi_B(x_1,b_1)\phi_{M_2}(x_2)\phi_{M_3}(x_3)\nonumber \\
 & &\times r_3 \left[-x_3 E_b(t_{en}^{(1)})a_i(t_{en}^{(1)})h^{(1)}_{en}(x_i,b_i)
 +\big(1+x_3+r_2(2x_2-2x_3+1)\big)E_{en}(t_{en}^{(2)})a_i(t_{en}^{(2)})h^{(2)}_{en}(x_i,b_i) \right]\;,
 \end{eqnarray}
 \begin{eqnarray}
 F_{en}^{LL,p}(a_i(t))&=& 16\pi\sqrt{\frac{2}{3}} C_F m_B^4 \int_0^1
 [dx]\int_0^{1/\Lambda} b_1 db_1 b_3 db_3
 \phi_B(x_1,b_1)\phi_{M_2}(x_2)\phi_{M_3}(x_3)\nonumber \\
 & &\times r_3 \left[-x_3E_b(t_{en}^{(1)})a_i(t_{en}^{(1)})h^{(1)}_{en}(x_i,b_i)
 +(1+x_3-r_2)E_{en}(t_{en}^{(2)})a_i(t_{en}^{(2)})h^{(2)}_{en}(x_i,b_i) \right]\;,
 \end{eqnarray}
 \begin{eqnarray}
 F_{en}^{LR,s}(a_i(t))&=& 16\pi\sqrt{\frac{2}{3}} C_F m_B^4 \int_0^1
 [dx]\int_0^{1/\Lambda} b_1 db_1 b_3 db_3
 \phi_B(x_1,b_1)\phi_{M_2}(x_2)\phi_{M_3}(x_3)\nonumber \\
 & &\times \left[-\big(x_2r_2^2-x_2r_2+r_3^2x_3\big)E_{en}(t_{en}^{(1)})a_i(t_{en}^{(1)})h^{(1)}_{en}(x_i,b_i)\right.\nonumber\\
 &&\left.-\big(x_2r_2^2-x_2r_2-r_3^2x_3\big)E_{en}(t_{en}^{(2)})a_i(t_{en}^{(2)})h^{(2)}_{en}(x_i,b_i) \right]\;,
 \end{eqnarray}
 \begin{eqnarray}
 F_{en}^{LR,p}(a_i(t))&=& 16\pi\sqrt{\frac{2}{3}} C_F m_B^4 \int_0^1
 [dx]\int_0^{1/\Lambda} b_1 db_1 b_3 db_3
 \phi_B(x_1,b_1)\phi_{M_2}(x_2)\phi_{M_3}(x_3)\nonumber \\
 & &\times \left[-\big(x_2r_2^2-x_2r_2-r_3^2x_3\big) E_{en}(t_{en}^{(1)})a_i(t_{en}^{(1)})h^{(1)}_{en}(x_i,b_i)\right.\nonumber\\
 &&\left.-\big(x_2r_2^2-x_2r_2+r_3^2x_3\big) E_{en}(t_{en}^{(2)})a_i(t_{en}^{(2)})h^{(2)}_{en}(x_i,b_i) \right]\;,
 \end{eqnarray}
 \begin{eqnarray}
 F_a^{LL,s}(a_i(t))&=&8\pi C_Ff_B m_B^4 \int_0^1
 dx_2dx_3\int_0^{1/\Lambda}b_2db_2b_3db_3 \phi_{M_2}(x_2) \phi_{M_3}(x_3)\nonumber \\
 & &\times r_2r_3 \left[(2-x_2)E_a(t_a^{(1)})a_i(t_a^{(1)}) h_a(1-(1-r_2^2)x_3,1-(1-r_3^2)x_2,b_3,b_2)S_t(x_2)\right.\nonumber \\
 & &\left.+(x_3-2)E_a(t_a^{(2)})a_i(t_a^{(2)}) h_a(1-(1-r_3^2)x_2,1-(1-r_2^2)x_3,b_2,b_3)S_t(x_3)\right]\;,
 \end{eqnarray}
 \begin{eqnarray}
 F_a^{LL,p}(a_i(t))&=&8\pi C_Ff_B m_B^4 \int_0^1
 dx_2dx_3\int_0^{1/\Lambda}b_2db_2b_3db_3 \phi_{M_2}(x_2) \phi_{M_3}(x_3)\nonumber \\
 & &\times r_2r_3\left[x_2E_a(t_a^{(1)})a_i(t_a^{(1)}) h_a(1-(1-r_2^2)x_3,1-(1-r_3^2)x_2,b_3,b_2)S_t(x_2)\right.\nonumber \\
 & &\left.+x_3E_a(t_a^{(2)})a_i(t_a^{(2)}) h_a(1-(1-r_3^2)x_2,1-(1-r_2^2)x_3,b_2,b_3)S_t(x_3)\right]\;,\\
 F_a^{LR,s}(a_i(t))&=&F_a^{LL,s}(a_i(t)),\\
 F_a^{LR,p}(a_i(t))&=&-F_a^{LL,p}(a_i(t)),
 \end{eqnarray}
 \begin{eqnarray}
 F_a^{SP,s}(a_i(t))&=&16\pi C_Ff_B m_B^4 \int_0^1
 dx_2dx_3\int_0^{1/\Lambda}b_2db_2b_3db_3 \phi_{M_2}(x_2) \phi_{M_3}(x_3)\nonumber \\
 & &\times \left[-r_3E_a(t_a^{(1)})a_i(t_a^{(1)}) h_a(1-(1-r_2^2)x_3,1-(1-r_3^2)x_2,b_3,b_2)S_t(x_2)\right.\nonumber \\
 & &\left.-r_2E_a(t_a^{(2)})a_i(t_a^{(2)})
 h_a(1-(1-r_3^2)x_2,1-(1-r_2^2)x_3,b_2,b_3)S_t(x_3)\right]\;,\\
 F_a^{SP,p}(a_i(t))&=&F_a^{SP,s}(a_i(t)),
\end{eqnarray}
\begin{eqnarray}
 F_{ac}^{LL,s}(a_i(t))&=&8\pi C_Ff_B m_B^4 \int_0^1
 dx_2dx_3\int_0^{1/\Lambda}b_2db_2b_3db_3 \phi_{M_2}(x_2) \phi_{M_3}(x_3)\nonumber \\
 & &\times \left[-r_2\big(r_2-r_3(x_3+1)\big)E_a(t_a^{(1c)})a_i(t_a^{(1c)}) h_a(x_2,x_3(1-r_2^2-r_3^2),b_2,b_3)S_t(x_3)\right.\nonumber \\
 & &\left.+r_3\big(r_3-r_2(x_2+1)\big) E_a(t_a^{(2c)})a_i(t_a^{(2c)}) h_a(x_3,x_2(1-r_2^2-r_3^2),b_3,b_2)S_t(x_2)\right]\;,
 \end{eqnarray}
 \begin{eqnarray}
 F_{ac}^{LL,p}(a_i(t))&=&8\pi C_Ff_B m_B^4 \int_0^1
 dx_2dx_3\int_0^{1/\Lambda}b_2db_2b_3db_3 \phi_{M_2}(x_2) \phi_{M_3}(x_3)\nonumber \\
 & &\times \left[r_2\big(r_2-r_3(x_3-1)\big)E_a(t_a^{(1c)})a_i(t_a^{(1c)}) h_a(x_2,x_3(1-r_2^2-r_3^2),b_2,b_3)S_t(x_3)\right.\nonumber \\
 & &\left.+r_3\big(r_3-r_2(x_2-1)\big) E_a(t_a^{(2c)})a_i(t_a^{(2c)}) h_a(x_3,x_2(1-r_2^2-r_3^2),b_3,b_2)S_t(x_2)\right]\;,
 \end{eqnarray}
 \begin{eqnarray}
 F_{ac}^{LR,s}(a_i(t))&=&F_{ac}^{LL,s}(a_i(t)),\\
 F_{ac}^{LR,p}(a_i(t))&=&-F_{ac}^{LL,p}(a_i(t)),\\
 F_{an}^{LL,s}(a_i(t))&=& 16\pi\sqrt{\frac{2}{3}} C_F m_B^4 \int_0^1
 [dx]\int_0^{1/\Lambda}b_1 db_1 b_2db_2\phi_B(x_1,b_1)\phi_{M_2}(x_2)\phi_{M_3}(x_3) \nonumber \\
 & &\times \left[-\big(x_2r_2^2-2r_2r_3+r_3^2x_3\big) E_{an}(t_{an}^{(1)})a_i(t_{an}^{(1)})h^{(1)}_{an}(x_i,b_i)\right.\nonumber\\
 &&\left.+ \big((x_2-1)r_2^2+r_3^2(x_3-1)\big)E_{an}(t_{an}^{(2)})a_i(t_{an}^{(2)})h^{(2)}_{an}(x_i,b_i) \right]\;,
 \end{eqnarray}
 \begin{eqnarray}
 F_{an}^{LL,p}(a_i(t))&=& 16\pi\sqrt{\frac{2}{3}} C_F m_B^4 \int_0^1
 [dx]\int_0^{1/\Lambda}b_1 db_1 b_2db_2\phi_B(x_1,b_1)\phi_{M_2}(x_2)\phi_{M_3}(x_3) \nonumber \\
 & &\times \left[\big(r_3^2x_3-r_2^2x_2\big) E_{an}(t_{an}^{(1)})a_i(t_{an}^{(1)})h^{(1)}_{an}(x_i,b_i)\right.\nonumber\\
 &&\left.+ \big(r_2^2(x_2-1)-r_3^2(x_3-1)\big)E_{an}(t_{an}^{(2)})a_i(t_{an}^{(2)})h^{(2)}_{an}(x_i,b_i) \right]\;,
 \end{eqnarray}
 \begin{eqnarray}
 F_{an}^{LR,s}(a_i(t))&=& 16\pi\sqrt{\frac{2}{3}} C_F m_B^4 \int_0^1
 [dx]\int_0^{1/\Lambda}b_1 db_1 b_2db_2\phi_B(x_1,b_1)\phi_{M_2}(x_2)\phi_{M_3}(x_3) \nonumber \\
 & &\times \left[\big(r_2(x_2+1)-r_3(x_3+1)\big) E_{an}(t_{an}^{(1)})a_i(t_{an}^{(1)})h^{(1)}_{an}(x_i,b_i)\right.\nonumber\\
 &&\left.-\big(r_2(x_2-1)-r_3(x_3-1)\big) E_{an}(t_{an}^{(2)})a_i(t_{an}^{(2)})h^{(2)}_{an}(x_i,b_i) \right]\;,
 \end{eqnarray}
 \begin{eqnarray}
 F_{an}^{LR,p}(a_i(t))&=&F_{an}^{LR,s}(a_i(t)),\\
 F_{an}^{SP,s}(a_i(t))&=& 16\pi\sqrt{\frac{2}{3}} C_F m_B^4 \int_0^1
 [dx]\int_0^{1/\Lambda}b_1 db_1 b_2db_2\phi_B(x_1,b_1)\phi_{M_2}(x_2)\phi_{M_3}(x_3) \nonumber \\
 & &\times \left[-\big(x_2r_2^2-2r_2r_3+r_3^2x_3\big) E_{an}(t_{an}^{(1)})a_i(t_{an}^{(1)})h^{(1)}_{an}(x_i,b_i)\right.\nonumber\\
 &&\left.+\big((x_2-1)r_2^2+r_3^2(x_3-1)\big) E_{an}(t_{an}^{(2)})a_i(t_{an}^{(2)})h^{(2)}_{an}(x_i,b_i) \right]\;,
 \end{eqnarray}
 \begin{eqnarray}
 F_{an}^{SP,p}(a_i(t))&=& 16\pi\sqrt{\frac{2}{3}} C_F m_B^4 \int_0^1
 [dx]\int_0^{1/\Lambda}b_1 db_1 b_2db_2\phi_B(x_1,b_1)\phi_{M_2}(x_2)\phi_{M_3}(x_3) \nonumber \\
 & &\times \left[\big(r_2^2x_2-r_3^2x_3\big) E_{an}(t_{an}^{(1)})a_i(t_{an}^{(1)})h^{(1)}_{an}(x_i,b_i)\right.\nonumber\\
 &&\left.-\big(r_2^2(x_2-1)-r_3^2(x_3-1)\big) E_{an}(t_{an}^{(2)})a_i(t_{an}^{(2)})h^{(2)}_{an}(x_i,b_i) \right]\;,
 \end{eqnarray}
 \begin{eqnarray}
 F_{anc}^{LL,s}(a_i(t))&=& 16\pi\sqrt{\frac{2}{3}} C_F m_B^4 \int_0^1
 [dx]\int_0^{1/\Lambda}b_1 db_1 b_2db_2\phi_B(x_1,b_1)\phi_{M_2}(x_2)\phi_{M_3}(x_3) \nonumber \\
 & &\times \left[\big((x_2-1)r_2^2+2r_2r_3+r_3^2(x_3-1)\big) E_{an}(t_{an}^{(1c)})a_i(t_{an}^{(1c)})h^{(1c)}_{an}(x_i,b_i)\right.\nonumber\\
 &&\left.-\big(x_2r_2^2+x_3r_3^2\big) E_{an}(t_{an}^{(2c)})a_i(t_{an}^{(2c)})h^{(2c)}_{an}(x_i,b_i) \right]\;,
 \end{eqnarray}
 \begin{eqnarray}
 F_{anc}^{LL,p}(a_i(t))&=& 16\pi\sqrt{\frac{2}{3}} C_F m_B^4 \int_0^1
 [dx]\int_0^{1/\Lambda}b_1 db_1 b_2db_2\phi_B(x_1,b_1)\phi_{M_2}(x_2)\phi_{M_3}(x_3) \nonumber \\
 & &\times \left[-\big(r_2^2(x_2-1)-r_3^2(x_3-1)\big) E_{an}(t_{an}^{(1c)})a_i(t_{an}^{(1c)})h^{(1c)}_{an}(x_i,b_i)\right.\nonumber\\
 &&\left.+\big(x_2r_2^2-x_3r_3^2\big) E_{an}(t_{an}^{(2c)})a_i(t_{an}^{(2c)})h^{(2c)}_{an}(x_i,b_i) \right]\;,
\end{eqnarray}
\begin{eqnarray}
 F_{anc}^{SP,s}(a_i(t))&=& 16\pi\sqrt{\frac{2}{3}} C_F m_B^4 \int_0^1
 [dx]\int_0^{1/\Lambda}b_1 db_1 b_2db_2\phi_B(x_1,b_1)\phi_{M_2}(x_2)\phi_{M_3}(x_3) \nonumber \\
 & &\times \left[\big((x_2-1)r_2^2+2r_2r_3+r_3^2(x_3-1)\big) E_{an}(t_{an}^{(1c)})a_i(t_{an}^{(1c)})h^{(1c)}_{an}(x_i,b_i)\right.\nonumber\\
 &&\left.+\big(r_2^2(x_2-1)-r_3^2(x_3-1)\big) E_{an}(t_{an}^{(2c)})a_i(t_{an}^{(2c)})h^{(2c)}_{an}(x_i,b_i) \right]\;,
 \end{eqnarray}
 \begin{eqnarray}
 F_{anc}^{SP,p}(a_i(t))&=&  16\pi\sqrt{\frac{2}{3}} C_F m_B^4 \int_0^1
 [dx]\int_0^{1/\Lambda}b_1 db_1 b_2db_2\phi_B(x_1,b_1)\phi_{M_2}(x_2)\phi_{M_3}(x_3) \nonumber \\
 & &\times \left[\big(r_2^2(x_2-1)-r_3^2(x_3-1)\big) E_{an}(t_{an}^{(1c)})a_i(t_{an}^{(1c)})h^{(1c)}_{an}(x_i,b_i)\right.\nonumber\\
 &&\left.-\big(r_2^2x_2-r_3^2x_3\big) E_{an}(t_{an}^{(2c)})a_i(t_{an}^{(2c)})h^{(2c)}_{an}(x_i,b_i) \right]\;,
\end{eqnarray}

\end{appendix}


\end{document}